\documentclass[longauth]{aa}

\usepackage{comment}
\usepackage{graphicx} 
\usepackage[normalem]{ulem} 
\usepackage[varg]{txfonts}
\usepackage{amssymb}
\usepackage{mathrsfs}
\usepackage{enumerate}
\usepackage{subfigure}
\usepackage{multirow}
\usepackage{siunitx}
\usepackage[colorlinks=true,allcolors=blue]{hyperref}%
\usepackage[dvipsnames]{xcolor}%
\usepackage{siunitx}
\usepackage{ulem}

\renewcommand*\vec[1]{\ensuremath{\mathbf{#1}}}

 \newcommand{\sect}[1]{Sect.~\ref{sec:#1}}

\defcitealias{ESTSRD}{EST SRD}

\hypersetup{colorlinks=true,citecolor=cyan,urlcolor=cyan,linkcolor=blue}

\begin{document}


\title{White paper on the relevance of the European Solar Telescope (EST) for the French 
heliophysics 
community.}
 
\titlerunning{EST \& the French heliophysics community}
\authorrunning{Pariat et al.}

\author{
 Contributing authors: \\
E. Pariat\inst{1,2}\orcid{0000-0002-2900-0608}, 
Q. Noraz\inst{3} \orcid{0000-0002-7422-1127},
B. Perri\inst{4}, \orcid{0000-0002-2137-2896},
N. Poirier\inst{5} \orcid{0000-0002-1814-4673}, 
C. Froment\inst{5} \orcid{0000-0001-5315-2890},
L. Bigot\inst{6}, \orcid{0009-0004-9005-1481},
G. Aulanier\inst{2} \orcid{0000-0001-5810-1566},
\& B. Gelly\inst{1} \orcid{0009-0002-0792-345X}
\\
Co-signing authors: \\
J. Aboudarham\inst{7} \orcid{0000-0002-0156-8162},
S. Aizawa\inst{2} \orcid{0000-0002-3483-3444},
O. Alexandrova\inst{7} \orcid{0000-0003-3811-2991},
S. Alqeeq\inst{2} \orcid{0000-0002-5401-6644},
T. Amari\inst{8} \orcid{0000-0002-2115-1284},
F. Auchère\inst{9} \orcid{0000-0003-0972-7022},
G. Bernoux\inst{10} \orcid{0000-0002-6823-4407},\\
M. Berthomier\inst{2} \orcid{0000-0001-6235-5382},
V. Bommier\inst{7} \orcid{0000-0002-6253-9170},
X. Bonnin\inst{7} \orcid{0000-0003-4217-7333},
P. Boumier\inst{9},
A. S. Brun\inst{4} \orcid{0000-0002-1729-8267},
I. Bualé\inst{7},
E. Buchlin\inst{9} \orcid{0000-0003-4290-1897},
A. Canou\inst{8},	\\
P. Canu\inst{2} \orcid{0000-0002-2517-4846},
T. Corbard\inst{6} \orcid{0000-0002-9615-9619},
F. Cornu\inst{7},
C. Coustillet\inst{11},
G. Cozzani\inst{5} \orcid{0000-0001-5515-1998},
L. D’herbomez\inst{2} \orcid{0009-0002-6403-6603},
S. Diaz Castillo\inst{1,12} \orcid{0000-0002-5330-3131},\\
T. Dudok de Wit\inst{5,13} \orcid{0000-0002-4401-0943},
M. Faurobert\inst{6} \orcid{0000-0001-9728-1782},
A. Finley\inst{14} \orcid{0000-0002-3020-9409},
D. Fontaine\inst{2} \orcid{0000-0002-4348-2623},
R. A. Garcia\inst{4} \orcid{0000-0002-8854-3776},
R. Grappin\inst{2} \orcid{0000-0001-7847-3586},
L. Hadid\inst{2} \orcid{0000-0002-8587-0202},\\
K. H. Henadhira Arachchige\inst{4},
M. Janvier\inst{9,14} \orcid{0000-0002-6203-5239},
L. Jouve\inst{15} \orcid{0000-0003-0657-4737},
R. Kieokaew	\inst{15} \orcid{0000-0003-0937-2655},
H. Kirkwood \inst{5},
B. Lavraud	\inst{16} \orcid{0000-0001-6807-8494},\\
J.-P. Le Breton\inst{5},
O. Le Contel\inst{2} \orcid{0000-0003-2713-7966},
N. Le Nestour\inst{5}, 
F. Leblanc\inst{17} \orcid{0000-0002-5548-3519},
S. Masson\inst{2,18} \orcid{0000-0002-6376-1144},
N. Meyer-Vernet\inst{8} \orcid{0000-0001-6449-5274},
S. Parenti\inst{4} \orcid{0000-0003-1438-1310},\\
F. Pitout\inst{15} \orcid{0000-0002-7073-4515},
M. Rieutord\inst{15} \orcid{0000-0002-9395-6954}, 
J. Romero Castañeda\inst{2,1} \orcid{0009-0000-2243-5302},
A. Rouillard\inst{15} \orcid{0000-0003-4039-5767},
C. Ruiz de Galarreta \inst{19,9},
F. Sahraoui\inst{2} \orcid{0000-0002-5973-8339},\\
B. Schmieder\inst{7} \orcid{0000-0003-3364-9183},
P. Simon\inst{2} \orcid{0000-0003-2091-2093}, 
A. Strugarek\inst{4} \orcid{0000-0002-9630-6463},
M. Tallon \inst{20} \orcid{0000-0002-6996-580X}
P. Thepthong \inst{5} \orcid{0000-0001-9597-1448},
J. Touresse\inst{2} \orcid{0009-0003-5268-5128},
J.-C. Vial\inst{9} \orcid{0000-0002-1511-9835},  \\
N. Vilmer\inst{7} \orcid{0000-0002-6872-3630},
\& A. Zaslavsky\inst{7} \orcid{0000-0001-8543-9431}\\
\vspace{0.2cm}
\email{etienne.pariat@themis.iac.es}
}
\institute{French-Spanish Laboratory for Astrophysics in Canarias (FSLAC), CNRS IRL2009, Instituto de Astrofísica de Canarias, La Laguna, Tenerife, France \& Spain  
 \and Laboratoire de Physique des Plasmas (LPP), CNRS, Ecole polytechnique, IPP, Sorbonne Université, Université Paris-Saclay, Observatoire de Paris-PSL, Palaiseau, France
 \and Centre for Mathematical Plasma-Astrophysics, Department of Mathematics, KU Leuven, Leuven, Belgium 
 \and Département d'Astrophysique/AIM CEA/IRFU, CNRS/INSU, Université Paris-Saclay, Université de Paris Cité, Gif-sur-Yvette, France 
 \and Laboratoire de Physique et de Chimie de l'Environnement et de l'Espace (LPC2E), OSUC, Université d'Orleans, CNRS, CNES, Orleans, France 
 \and Université Côte d'Azur, Observatoire de la Côte d'Azur, CNRS, Lagrange, Nice, France 
  \and Laboratoire d’Instrumentation et de Recherche en Astrophysique (LIRA), Observatoire de Paris-PSL, CNRS, Sorbonne Université, Université Paris-Cité, Meudon, France 
 \and CNRS, Centre de Physique Théorique de l'Ecole Polytechnique (CPHT), IPP, Palaiseau, France 
 \and Institut d'Astrophysique Spatiale (IAS), CNRS, Université Paris-Sud, Université Paris-Saclay, Orsay, France 
 \and Département Physique, instrumentation, environnement, espace (DPHY), ONERA, Université de Toulouse, Toulouse, France 
 \and Institut Supérieur de l’Aéronautique et de l’Espace (ISAE-SUPAERO), Université de Toulouse, Toulouse, France 
 \and Institut für Sonnenphysik (KIS), Freiburg, Germany 
 \and International Space Science Institute (ISSI), Bern, Switzerland 
 \and ESTEC - European Space Agency, Noordwijk, The Netherlands 
 \and Institut de Recherche en Astrophysique et Planétologie (IRAP), Université de Toulouse – Observatoire Midi-Pyrénées, Toulouse, France 
 \and Laboratoire d'Astrophysique de Bordeaux (LAB), Université de Bordeaux, CNRS, Pessac, France 
 \and Laboratoire Atmosphères, Observations Spatiales (LATMOS), CNRS, Sorbonne Université, Paris, France 
 \and Observatoire radioastronomique de Nancay (ORN), Observatoire de Paris-PSL, CNRS, Université d'Orléans, Nancay, France 
\and Instituto de Astrofísica de Canarias, 38205 La Laguna, Tenerife, Spain  
\and Centre de recherche astrophysique de Lyon (CRAL), Université Claude Bernard Lyon 1, CNRS, ENS, Saint Genis Laval, France 
}

\date{Public release of version 1: March 30th, 2026}
\abstract
{The project of the European Solar Telescope aims to provide a state-of-the art infrastructure to study the Sun and its interactions with Earth and the heliosphere. This $4.2$m aperture telescope will be equipped with multi conjugate adaptive optics, light-polarisation analyser, imaging spectrograph and integral field unit spectrographs. It will provide unprecedented observations of the solar photosphere and chromosphere and of the dynamical events and features that pertains to the low solar atmosphere. The EST project is presently in a phase of crystallisation, aiming at the creation of an European Research Infrastructure Consortium. While the French community has continuously been associated with the development of the EST project, some specific scientific aspects are more particularly relevant for the French astrophysics and heliophysics communities. The present review highlights the scientific research axes of high interest from the French community that shall strongly benefit from EST. The later will not only advance numerous topics of solar physics, as well as solar adaptive optics developments, but will also provide unrivaled datasets of high interest in the framework of space weather. This review also aims to highlight the space weather use that can be done with future EST observations, that will be particularly relevant for French heliophysicists.}


\maketitle

\section{Introduction: the EST project and the objectives of this white paper} \label{sec:Introduction}

\subsection{The EST project}

 The \href{https://est-east.eu/}{European Solar Telescope (EST)}\footnote{\url{https://est-east.eu/}} 
 is the unique European project of the next-generation large-aperture solar telescope \citep{Collados08,Collados10,Collados13}. With a $4.2$m primary mirror, it will be optimised for studies of the magnetic coupling of the different layers of the solar atmosphere, from the deep
photosphere to the upper chromosphere \citep[see][the reference scientific paper on EST]{Noda22}. EST will be located in La Palma, in the Canary Islands, a first-class site for astronomical observations. EST shall become the most powerful European ground-based facility to study the Sun in the next 2-3 decades, in the visible and near-infrared bands. EST shall both complement and compete with the US Daniel K. Inoue Solar Telescope \citep[DKIST, ][]{Rimmele20}, the unique existing 4m-class solar telescope EST will provide diagnostics of the thermal, dynamic and magnetic properties of the plasma over many scale heights, by using multi-wavelength imaging, spectroscopy and spectropolarimetry. Thanks to its multi-conjugate adaptive optics (MCAO) and polarisation analysis capacities, EST will specialise in high-spatial and temporal resolution, using several instruments simultaneously to efficiently produce 2D spectral information in different wavelength domains with complementary methodology (imaging spectrograph, integral field unit spectrographs). 
 
The major science questions that drive the EST project are: 
\begin{itemize}
\item how does the magnetic field emerges to the surface and evolves? 
\item how is the energy  transported from the photosphere to the chromosphere?
\item how is the energy released and deposited in the upper
atmosphere?
\item how are the major solar eruptions, possibly affecting human activities, induced?
\end{itemize}
Examples of major scientific themes of study by EST are quiet Sun magnetism and its impact on the chromosphere and upper layers, magneto-acoustic and Alfvén waves, chromospheric heating and mass loading, localised reconnection
events, flares, filament eruptions, instabilities, non-ideal magnetohydrodynamics (MHD) effects, space weather inducing events. The reader is referred to the Section 3 of \citet{Noda22} for a comprehensive list of EST science topics, as well as to  the Science Requirement Document\footnote{\url{https://www.est-east.eu/science-requirement-document}} \citep[][hereafter \citetalias{ESTSRD}]{ESTSRD} written by the EST science advisory group. 
 
EST is now a mature project. Its Conceptual Design Study took place in 2008-2011 \citep{Collados08,Collados10}, funded by the FP7 program of the European Commission (FP7). This was followed by the "Getting Ready for EST" (GREST) project, in 2015–2018, funded through the H2020 program, which undertook activities to improve the performance of instrumentation.  In parallel, the two EU SOLARNET projects provided open access to first-class infrastructures, networking opportunities, and joint research and development activities between different European facilities. In 2016, the EST project was included in the European Roadmap for Research Infrastructures (ESFRI). The EST Preparatory Phase (PRE-EST) started after that milestone, producing detailed implementation plans, governance options, procurement strategies, site comparisons, risk analyses and detailed designs for key EST elements. 
EST is also an identified top priority ground-based infrastructure of ASTRONET\footnote{ \href{https://www.astronet-eu.org/?page_id=521
}{Astronet Roadmap 2022-2035 report}}, the consortium of European funding organisations and infrastructures. 

EST is now in an intermediary phase between the preparatory phase and the construction phase. In 2023, the European Solar Telescope Canarian Foundation (EST-CF) has been created to manage and steer the EST project, consolidating the funds for the construction of EST, making possible to start staff recruitment for construction activities as soon as construction funds are secured.  Currently, the EST-CF is formed by ten research institutions from eight European countries, with no French institution at this time. In 2024 \& 2025, the EST-CF has driven to success several Preliminary Design Reviews (PDR) of the EST subsystems, namely optics and adaptive optics, control system, adaptive optics real-time controller, data centre, project management, as well as the Conceptual design review (CDR) of the EST scientific instrument suite. The EST-CF targets the creation of an European Research Infrastructure Consortium (ERIC), which will have the legal status to start the effective construction of the EST infrastructure. As of today three countries are committed to the future ERIC: Spain, Slovakia and Czech Republic. These countries are engaging themselves to fund about $40\%$ of the $250-300$ M€ total budget of the EST project.  

\subsection{White paper objectives}

The EST project is thus presently at a crucial stage, situation that is driving the need of an in-depth reflection on the relevance and implication of the French Astrophysics community on the EST project. 

Thus, following a first workshop that occurred in Meudon in 2010, the \emph{EST France 2025 workshop}\footnote{\url{ https://est-france-2025.sciencesconf.org/}} took place in Paris in June 2025, with the objectives to bring together EST’s European leadership with French scientific and industrial communities interested in the project’s implementation. The workshop presented the EST project, its significance, and its potential uses for the French community, as well as facilitated discussions on France’s strategic involvement in the EST scientific consortium. Publicly available presentations given during the workshop can be accessed through a dedicated shared repertory\footnote{ \href{https://cloudsrv.themis.iac.es/nextcloud/index.php/s/JGn5t4xq5ZqBA7L}{EST France Workshop 2025 presentation repository}  (under Creative Commons CC BY-NC 4.0 license)}. The workshop round-table discussions highlighted the need to write the present white paper on the renewed strong interest of the French community for the EST project. 

\newcounter{Savefootnotenum}
In its review and outlook report, "Rapport de bilan 2019-2023 et prospective 2025-2029 du PNST"\footnote{ \href{ https://atst.osups.universite-paris-saclay.fr/wp-content/uploads/2020/01/Bilan_et_Prospective_PNST_2024_versionfinale18juin-cf0.pdf}{Full report (in French)}. See also its \href{ https://atst.osups.universite-paris-saclay.fr/wp-content/uploads/2020/01/Executive_Summary_prospective_PNST_2024_Final-02e.pdf}{executive summary (in French)}}, 
\setcounter{Savefootnotenum}{\value{footnote}}
the "Action Thématique Soleil-Terre"\footnote{\url{https://atst.osups.universite-paris-saclay.fr/}}, ATST, (formally known as the Programme National Soleil-Terre, PNST), which gathers the whole French heliophysics community, reports the EST project as a leading priority of ground-based projects. It states: {\it The solar community is no longer in a position to provide a Principal Investigator for an instrumental contribution to this ground-based project, but remains highly interested in the science and in international cooperation around EST.} Meanwhile, in the latest roadmap\footnote{\url{https://prospective-aa.sciencesconf.org/}}
 of the research community of the Astronomy and Astrophysics domain of the Institut National des Sciences de l’Univers (INSU) of CNRS, the EST project was not received as a priority. The report of the working group dedicated to priority projects \footnote{\href{https://prospective-aa.sciencesconf.org/data/rapport_synthese_groupeIII1_moyens_prioritaires_astro_2024.pdf}{Rapport de synthèse des travaux du groupe III.1: Moyens
prioritaires}} states: {\it Despite its scientific interest, the group reiterates the observation made during the last foresight exercise regarding the limited engagement of the French community in the preparatory studies (both technical and scientific) for the European Solar Telescope. This level of activity does not justify investing resources at the required scale.} The authors of the present white papers believe that these diverging views (of the ATST and of the INSU prospective group) may stems both from the lack of clear and explicit argumentation on the relevance of the EST project for the actual French astrophysics community, first "raison d'être" of the present document, as well as from the evolution of the scientific interests of the French community that may not have been fully perceived. 

Indeed, since the start of the EST Conceptual Design Study, which started in February 2008, almost 20 years ago, the landscape of French solar physics research has changed. Ground-based solar physics instrumentation was a strength of the French community at the dusk of the XX\textsuperscript{th} century, which drove the construction of the THEMIS solar telescope \citep{Rayrole85}. With the evolution of human resources, 
this branch has now been replaced by other roles on which the French solar community has an international leadership.
Thanks to the high level of structuring of the French community enabled by the the "Service Nationaux d'Observations"\footnote{\url{https://www.insu.cnrs.fr/fr/les-services-nationaux-dobservation}}
(SNO, national observation services) of INSU (e.g., 3SOLEIL, CDPP, CLIMSO, MEDOC, STORMS, etc), the creation of the French Organisation for Applied Research in Space Weather\footnote{\url{http://www.meteo-espace.fr/}} (OFRAME), the hiring of several French civil servants researchers with a strong focus toward space weather research, space weather research and monitoring is now a strength of the French heliophysics community.
On this aspect, the interested reader shall examine the SWOT analysis in the latest review and outlook report \footnotemark[\value{Savefootnotenum}] of ATST, and in particular the Sections 4, 5 $\&$ 8 of that report.
A new generation of solar physicists is ready to be involved, with new unique expertise on theory, modelling and data analysis at the interface of instrumentation. Moreover, the atmospheric layers of the Sun observed by EST are in connection between the solar interior and solar extended corona. The EST project can thus further connect and structure the French solar plasma community. The second "raison d'être" of this review is thus to introduce several space-weather and heliophysics science cases that EST is uniquely placed to address.

The objectives of the present white paper is thus to present to the interested reader an overview of the EST project research objectives which are specifically relevant to the French astrophysical community. The objectives are thus not to list all the research topics for which EST will be groundbreaking \citep[which are already largely listed in][]{Noda22}, but only those for which the French community can either contribute to the EST project, or the EST expected data products that are directly relevant to the development of innovative research axes of the French community.

This white paper is thus organised as follow: in Section \ref{sec:past} a brief summary of the past involvement of the French community in the EST project will be presented. Then, Sections  \ref{sec:solar} 
$\&$ \ref{sec:spaceweather}  
will illustrate science topics, respectively in  solar physics, 
and space weather research
, of interest for diverse French research teams. Finally in Section \ref{sec:conclusion}, an overview of the desired road-map for the French involvement in the EST project will be sketched.


\section{Summary of the past involvement of the French community on the EST project} \label{sec:past}

French teams and scientists have continuously been involved in the preparation of the EST project and its discussion. This factually took the form of French institutions being partner on different European Union funded project, as well as proposal for French-led instruments. 

The European Association for Solar Telescopes (EAST) was founded in 2006 by a group of research institutions from 14 European countries. Four more countries have joined since then. The goal of EAST is to ensure access of European solar astronomers to world-class high-resolution ground-based observing facilities. In order to achieve that goal, EAST has been promoting the development, construction and operation of EST. Since its creation, France has been continuously represented in EAST through the THEMIS team. 

France has been a force of proposition for instrumental developments on EST. The Multichannel Subtractive Double Pass (MSDP) spectrograph \citep{Mein77,Mein21} is a French solar instrument concept historically used at telescopes such as the Meudon Solar Tower,  Pic du Midi Turret Dome, VTT, THEMIS \& large Wroclaw coronagraph. \citet{Malherbe23a,Malherbe23b} proposed a MSDP to be incorporated to the EST visible and IR spectrographs lines, using new generation slicers (56 channels, high spectral resolution). The aim is such MSDP was to produce 56-channel spectra images with the spatial resolution of the AO and reconstitute cubes of instantaneous data $(X, Y, \lambda)$ at high cadence. However lack of perspective in PI-ship for this instrumental concept, that stems from the strong weakening of the French optical solar instrumentation community, has not permitted to reach sufficient momentum for MSDP to be considered as first-light instrument on EST.  

French teams and researchers have however been continuously involved with the EST project itself. Between 2008 and 2011, the conceptual design study of EST was realised as an EU FP7 project. It involved 14 research institutions and 15 industrial partners, among which IRAP/Université Paul Sabatier (UPS), the Paris Observatory and THEMIS \citep{Collados10}. The conceptual design phase was summarised in the “EST: Conceptual Design Study Report” containing the proposed solutions for the telescope itself and all its subsystems. B. Gelly was the workpackage leader of the optical design. A “Report on technical, financial, and socio-economic aspects” was also one of the conceptual phase project deliverables. 

The French community was not directly involved in GREST but participated to PRE-EST. Both IRAP and THEMIS teams have been involved in this consortium. PRE-EST’s Science requirement development relied on the European solar community’s experience with existing instruments. IRAP researchers, A. Lopez Ariste and K. Dalmasse, respectively participated to the EST science advisory group and to the redaction of \citetalias{ESTSRD}. French involvement reinforced EST science cases focused on chromospheric magnetism and precision polarimetry — helping set realistic performance targets and observational modes. 

In parallel, pan-european efforts to trains the next generation of european solar researchers, and connect tighter the different institutions to be involved in EST where carried through the two SOLARNET (High-Resolution Solar Physics Network) programs. Both 
included important French involvement. The first SOLARNET \footnote{\url{http://research.iac.es/proyecto/solarnet/}} aimed at bringing together and integrating the major European research infrastructures in the field of high-resolution solar physics in order to promote the future development of EST. The project was carried out between April 2013 and March 2017 and funded from the European Commission’s FP7 Capacities Programme. It involved 16 countries with French participation thorugh THEMIS/CNRS, IRAP/UPS as well as a private partner, Winlight Optics. France was also a partner in the second SOLARNET network\footnote{\url{https://www.solarnet-project.eu/}}, funded by the European Union's Horizon 2020 Research and Innovation Programme, that pursued training and coordination toward EST. It again involved 32 partners from 16 countries: France was represented by THEMIS/CNRS as a research institution and both ALPAO and Winlight Optics where among the 6 EU private companies involved in SOLARNET. This already demonstrate the strong French industrial interest in EST which complement French academic implication.

The French solar community is therefore explicitly referenced in the EST design-study materials as a contributor to the design/requirements and science case, with several french researchers, from almost all French laboratories in which solar physics research is carried (CRAL, GEPI, IRAP, LAGRANCE, LESIA - now LIRA -, LPP, THEMIS), co-signing the EST seminal paper \citep{Noda22}. As the following sections demonstrate, the French solar physics community is highly interested by the innovative science and the state-of-the-art data that EST shall provide, allowing a high scientific return for French activities in solar physics and space weather 
research.


\section{French solar physics research with EST} \label{sec:solar}

In this section, EST solar physics research axes of specific interest to the French community are presented. While they touch almost all main science questions of solar physics, the section focuses on the domain in which French solar physicists have provided valuable contribution in the past decades. Namely this concern the dynamics of granulation (Sec. \ref{sec:solar_convection}), and its underlying magnetism (Sec. \ref{sec:solar_QS}), the measurement of electric currents (Sec. \ref{sec:solar_electriccurrents}), the heating of the chromosphere (Sec. \ref{sec:solar_chromo}), the photosphere induced coronal oscillations (Sec. \ref{sec:solar_coronaloscillations}), coronal rain (Sec. \ref{sec:solar_coronalrain}), the signature of flares and eruptions (Sec. \ref{sec:solar_eruptions}), the solar wind helium abundance (Sec. \ref{sec:solar_helium_sw}), and the  extraction processes at the source of the solar wind (Sec. \ref{sec:solar_abundances}).

\subsection{Observation and modelling of the granulation scales}
\label{sec:solar_convection}

Underneath the photosphere lies the convective layer of the Sun, where the transfer of energy is done mostly by large-scale flows triggered by a difference in temperature between the hot bottom layers and the colder outer layers. Convection is a process that interacts with many key structures at the solar surface, and yet the exact nature of the convective scales is still an open question. Indeed, different observations yield different estimations of the toroidal kinetic spectra \citep{birchSolarConvectiveVelocities2024,Noraz2025}. As convection is a key process of the star evolution, this yields contradictory constraints for numerical simulations, that struggle to reproduce the solar differential rotation profile in the same range of parameters. This problem is refered to as the convective conundrum \citep{omaraVelocityAmplitudesGlobal2016,Hotta2023}. Hence, it is crucial to be able to better constraint the various convection scales, as it will help better constrain both local and global numerical simulations of the magnetic field generation and emergence.

\begin{figure}
    \centering
    \includegraphics[width=0.99\linewidth]{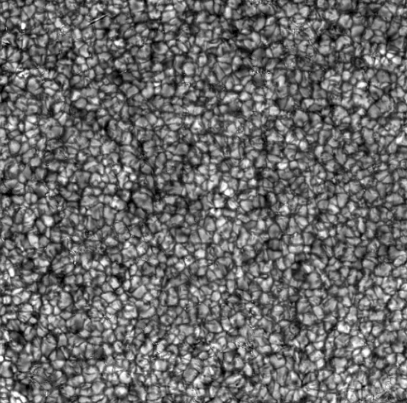}
    \caption{Example of solar granulation observed by THEMIS. From \cite{Roudier2020}.}
    \label{fig:granulation_convection}
\end{figure}

The french community has focused on studying convection from both an observational point of view \citep{Roudier2020,ZhengC24} and a numerical point of view \citep{Noraz2025}. These studies can help better characterize the convection through its most visible counterpart which is the solar granulation. 
Solar granulation represents the visible manifestation of convective energy transport in the Sun's photosphere. These cellular structures, typically $\sim$1000 km in size with lifetimes of 8--15 minutes, form the dominant pattern across the solar surface (cf. Figure~\ref{fig:granulation_convection}). Bright granules mark regions where hot plasma, with an excess enthalpy $\Delta H$, rises from the solar interior at velocities of $F_\odot/(\rho \Delta H) \approx$ 2 km/s, corresponding to $\sim$10--15\% of the sound speed \citep{nordlund85}. The darker intergranular lanes indicate cooler, descending material that has radiatively evacuated its excess enthalpy when reaching an optical depth of order unity \citep[and references therein]{nordlund09}. While the granulation pattern itself is not turbulence \citep{nordlund97}, the study of small-scale convective fluctuations within the granular pattern provides direct observational access to fundamental turbulence theory. Also, below the photosphere lies a bigger network of supergranules that can reach 30 000 km and live up to 24 hours. It is challenging to disentangle the dynamics of the granules from the supergranules. We can note that the dynamics of granules is also linked with the dynamics of the large-scale flow, and can help infer them with specific methods such as Coherent Structure Tracking \citep[CST ; ][]{Roudier2018}. 

On the simulation side, three-dimensional radiative hydrodynamical simulations of solar surface convection have been performed with remarkable realism since the pioneering work of Nordlund four decades ago \citep{nordlund82,nordlund85,stein98,nordlund09}, and are now routinely carried out by other groups \citep[and references therein]{rempel23}. The Quiet Sun surface is reproduced using Large Eddy Simulations (LES), in which the viscosity is purely artificial and defined according to the mesh size in order to ensure numerical stability. This approach assumes that the real viscosity values are so small that they do not affect the dynamics at the largest scales (i.e. granules). As a consequence, the power spectrum is truncated at sub-grid scales. The Prandtl number (ratio of viscosity to radiative diffusivity) in these simulations is far too large ($\sim$0.1--1.0) compared to its actual solar value ($\sim10^{-9}$). The success of the LES approach for stellar and solar granulation lies in the fact that the granulation scale is such that, fortunately, the P\'eclet number (ratio of advective to diffusive heat transport) is of order unity. As a result, heat transport can be realistically modeled in these LES despite the Reynolds and Rayleigh numbers being far from their true values. EST will provide measurements at scales well below the pressure scale height (the characteristic scale at the solar surface, $\sim$150 km), offering valuable information to explore the current sub-grid scales of LES and providing guidance for improving their modeling.

\begin{figure*}[ht!]
\centering
\includegraphics[width=0.37\linewidth]{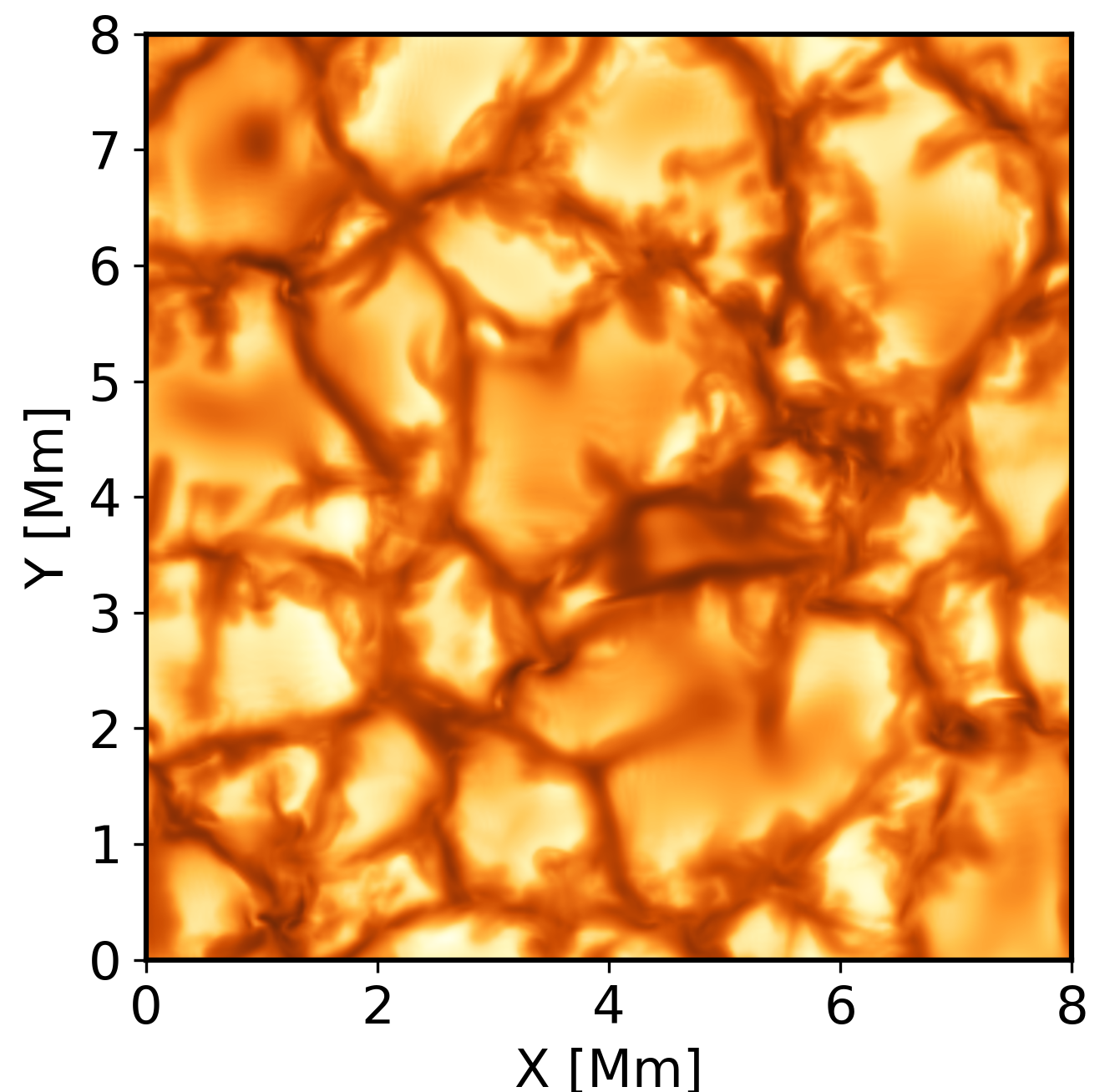}
\includegraphics[width=0.62\linewidth]{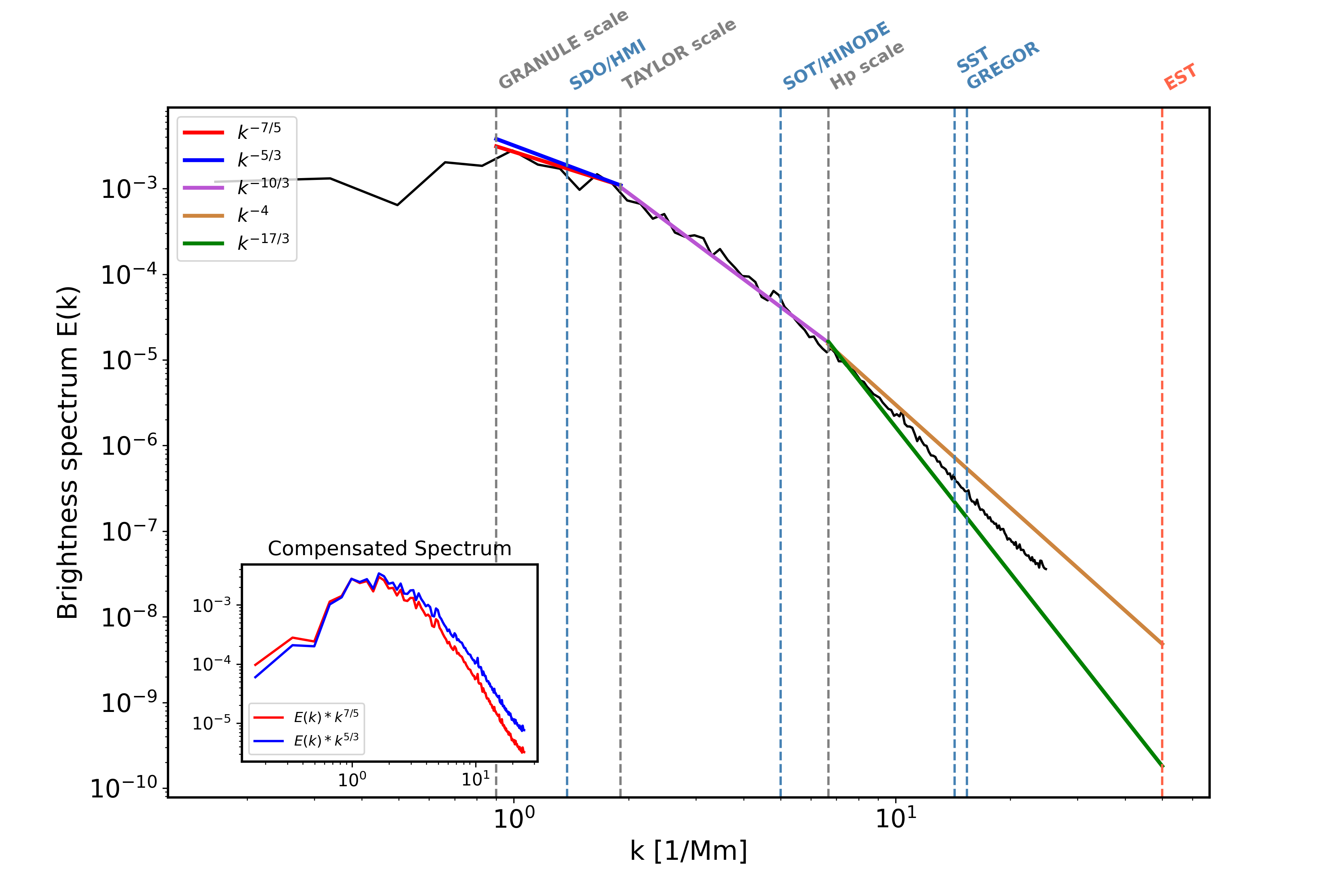}
\caption{
(Left panel) Disk-center emerging (bolometric) intensity from a radiation hydrodynamical simulation of solar granulation, performed with the \textit{Stagger-code} on an $8 \times 8$ Mm domain (25 km grid resolution). (Right panel) 
Power spectrum of the intensity as a function of wavenumber (1/Mm). Colored straight lines indicate turbulence regimes, and vertical lines indicate characteristic physical scales at the solar surface and the resolution limits of current solar telescopes (Bigot, in prep).}
\label{fig:quiet}
\end{figure*}

Such analyses, using solar brightness fluctuations to characterize the turbulence regime at the solar surface, have been performed in several studies \citep[e.g.][]{rieutord10} but they were limited due to the poor spatial resolution of the instruments. White-light space telescope \citep[e.g. HINODE][]{tsuneta08}, and 1-meter class solar telescopes, e.g. such as the Goode Solar Telescope \citep[GST][]{Goode12}, GREGOR \citep{schmidt12}, the Swedish Solar Telescope \citep[SST, see][]{Scharmer2019}, THEMIS \citep{Gelly16} have their spatial resolution at the solar surface limited to $\sim$70 km (i.e. only about half the pressure scale height),
whereas EST will reach $\sim$15 km. In this range of scales (15--150 km), the turbulence regime is expected to be the inertial--diffusive regime: velocity fluctuations remain dynamically active (inertial), while temperature gradients are increasingly smoothed by radiative diffusion \citep[e.g.][]{pope00}.

This behavior is illustrated in Figure~\ref{fig:quiet}, which shows the power spectrum of brightness fluctuations of solar granulation obtained from LES performed with the \textit{Stagger-code} \citep{stein24}. The power spectrum reveals the energy cascade from the injection range (granulation scales) down to the inertial--diffusive range. A classical Kolmogorov regime (i.e. assuming self-similarity of the turbulent cascade down to viscous dissipation scales) is not observed over all scales due to the strong stratification and the transition from optically thick to optically thin conditions at the solar surface \citep[e.g.][]{nordlund97,schumacher20}. While a quasi-Kolmogorov dependence ($k^{-5/3}$) is seen from the injection range down to the Taylor scale, the spectrum approaches the characteristic slope of the inertial--diffusive regime ($k^{-17/3}$) at the smallest scales. Clearly, the LES does not fully reproduce the expected slope at these scales because of the artificial viscosity employed. Future measurements with EST will provide critical observational constraints to calibrate the sub-grid regime of LES and will offer an exceptional testbed for turbulence theories in extreme conditions that are inaccessible in laboratory experiments ($P_r \ll 1$).

The high-resolution images of EST could provide a unique opportunity to better characterize the solar granulation, and hence the solar convection and the overall modeling of the quiet-sun. Such developments can have a strong impact on our ability to model the solar dynamo. 
The link between small and large convection and magnetic scales is essential to understand how the Sun organize its mean flows and cyclic 11-year dynamo \citep{Brun2017}. Being able to monitor changes in convective power, magnetic field organisation and complexity of flux emergence events along the 11-year cycle phase is key has it provides direct constraints on the underlying physical mechanism. 
Also, the subsurface layers of the Sun host strong rotational shear, as revealed by analyses of f-modes and ring-diagram helioseismology using MDI and HMI data \citep[e.g.][]{corbard02,cunnyngham17,komm22}. These studies find a shear $d\ln(\Omega)/d\ln r \approx -2$ to $-4$, depending on depth, latitude, and the phase of the solar cycle. Recently, \citet{faurobert23} and \citet{corbard25} introduced an innovative method based on measuring systematic shifts between simultaneous granulation images taken at different heights, using both the THEMIS and HINODE, to infer tiny rotation-induced displacements. The Fe~\textsc{i} 630.15 nm and Ca~\textsc{i} 616.2 nm lines were employed. While these measurements confirm a linear depth dependence of the rotation-rate gradient, the uncertainties remain large, limiting the study of solar-cycle variations. 
Local ring diagram or time-distance helioseismology techniques can further be used to put constraints on the solar inner dynamics and in particular on the study of torsional oscillations and how they may change depending on the level of solar activity \citep{Howe2013}. 
High-resolution EST observations of solar granulation will significantly reduce these uncertainties and will enable detailed studies of subsurface rotational flows and their implications for solar dynamo processes. 
This is also in agreement with the Section 4.9 of the \citetalias{ESTSRD}, which focuses on how the convection flows can erode sunspots and the surface magnetisme (see in particular Table 4.9.1 on combined spectropolarimetric observations with feature tracking).

\subsection{Magnetism of the Quiet-Sun}
\label{sec:solar_QS}

Solar granulation hosts the Quiet Sun magnetism. This small-scale turbulent magnetic field, known as the internetwork field, covers almost the entire solar surface \citep[and references therein]{bellot-rubio19}. Convective flows interact with magnetic fields emerging from deeper layers, creating complex couplings that drive phenomena ranging from small-scale magnetic flux emergence to the heating of the solar chromosphere and corona. While the origin of this Quiet Sun magnetism is not yet fully understood, it represents a vast magnetic reservoir whose total surface flux exceeds that of active regions. There is little doubt that this quiet magnetism affects the energy budget at the solar surface. Its role in chromospheric heating remains an active area of research and is discussed in Sec.~\ref{sec:solar_chromo}. Three-dimensional MHD simulations of the Quiet Sun suggest that this magnetic field arises from a small-scale dynamo, as theoretically predicted by \cite{kazantsev68,vincenzi02} in the limit of vanishing magnetic Prandtl number. Modern 3D MHD simulations \citep{voegler07,schussler08,rempel14} and recent reviews \citep{rempel23} incorporate physics compatible with solar surface conditions (e.g. compressibility, radiative transfer, and ionization). These simulations show that small-scale dynamo action can generate transverse magnetic fields of order $\sim$100 G, consistent with spectropolarimetric observations \citep{trujillo-bueno04}. However, these results should be interpreted with caution, as they are obtained with magnetic Prandtl numbers of order unity, which are many orders of magnitude larger than the true solar value \citep[e.g.][]{kapyla21}. Despite this limitation, small-scale dynamo simulations have been considered successful and have been used to argue that Quiet Sun magnetism is largely uncorrelated with the global solar magnetic field, since it is driven by random turbulent motions. This interpretation appears consistent with spectropolarimetric observations from the SOT/HINODE instrument \citep{tsuneta08}, which show no significant variation of Quiet Sun magnetism over the solar cycle \citep{lites08,lites14,faurobert15}.

However, this conclusion is not straightforward and has been challenged by several studies. For example, \cite{karak16} demonstrated numerically that, under certain conditions, small-scale and large-scale dynamos may be coupled rather than independent. Observational studies also suggest that the Quiet Sun exhibits solar-cycle-related variations. Using MDI/SOHO magnetograms, \cite{meunier18} found that Quiet Sun magnetic properties vary in phase with the solar cycle, while \cite{muller18} reported changes in granulation brightness contrast with increasing solar activity. More recently, \cite{Korpi-Lagg22} proposed that the apparent absence of solar-cycle variations in surface radial velocity measurements obtained with space-based spectrographs such as MDI \citep{scherrer95} or HMI \citep{scherrer12} is due to insufficient spatial resolution. Using the Fe~\textsc{i} 630.15 nm line observed with HINODE/SOT, \cite{faurobert16,faurobert21} also showed some changes with the solar cycle. Indeed, they showed that the temperature gradient varies over the solar cycle, indicating that the global magnetic field influences internetwork magnetism. They also reported a North--South hemispheric asymmetry. All these results, pointing to a solar-cycle dependence of Quiet Sun magnetism, have recently been confirmed using helioseismic data, in agreement with the observed $\sim$0.1\% variation of the total solar irradiance (Bigot, in prep.). Finally, using the granule tracking technique of on HMI/SDO data, \cite{ballot21} detected a small ($\sim$2\%) solar-cycle modulation of the average granule size, corresponding to $\sim$20 km, which is comparable to the expected spatial resolution of EST. The latter, thanks to its fine spatial resolution (more an order of magnitude better than HMI), will reveal the tiniest changes in granulation aspect. Combined with granule tracking or pattern-recognition techniques, EST will therefore enable precise measurements of granulation properties over the solar cycle. 
This is also in agreement with the Section 8.1 of the \citetalias{ESTSRD}, which focuses on small-scale magnetism in the quiet photophere (see in particular Tables 8.1.2 and 8.1.3 for the detection of polarisation fluctuations at the order of the granular scale).

\subsection{High-fidelity measurements of electric currents} \label{sec:solar_electriccurrents}

Magnetic fields represent the main source of energy of the major solar eruptive events. But the magnetic energy of solar active region is not necessarily readily available to be converted in kinetic energy and heating during solar flares and eruptions. A solar magnetic system which magnetic field is potential, i.e. which magnetic field can be describe as close to $\nabla \phi$,  can barely provide any of its energy \citep[][]{Schrijver05}. Only solar active systems that are highly non-potential, i.e. that posses a large amount of magnetic helicity, have "free" magnetic energy can induce significant eruptive behaviors. This non-potentiality is the direct consequence of the presence of intense electric currents. Current-carrying magnetic fields are an fundamental ingredient for the generation of flares and eruptions \citep[e.g.][]{Shibata11,Aulanier14}.  Several studies \citep[][]{Aulanier05,Aulanier06,Aulanier12,Janvier13,Janvier15},  have highlighted how electric currents sheets, and their low atmosphere signature, tightly relates with the very properties of the 3D magnetic reconnection occurring in the corona during eruptions (see Section \ref{sec:solar_eruptions}). In order to characterise the physical processes at work in a three-dimensional magnetic configuration before, during, and after eruptions, electric currents hence needs to be measured. 

\begin{figure}[ht!]
    \centering
   \includegraphics[width = 0.99\linewidth]{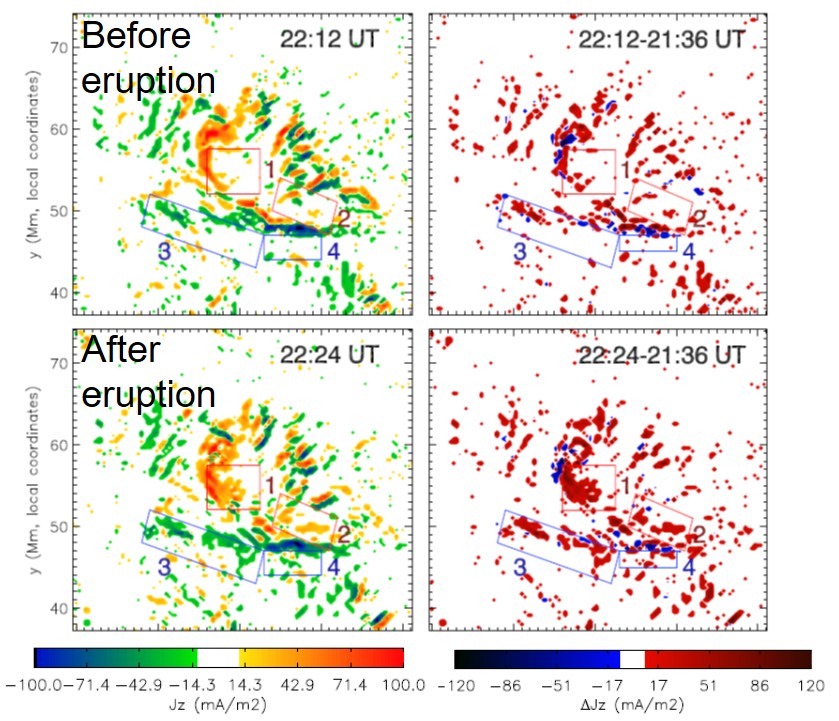}
    \caption{Distribution of the vertical component of the electric current density $j_z$ (left column), as well as their intensity difference (right column) before (top row) and after (bottom row) a major solar eruptions. Adapted from \citet{Janvier16}.}
    \label{fig:electriccurrents}
\end{figure}

Measuring electric currents and understanding their distribution \citep[e.g.][]{Bommier07}, has been a core theme of French solar physics research this past two decades. This is driven by their central role to describe and explain the structure of the coronal magnetic field.  Proper measure of the current allows to model the 3D coronal magnetic field  with a high level of fidelity, enabling precise comparison with eruption emission \citep[e.g.][]{Masson17}. Studying how current density distribution evolve during major solar eruptions, \citet{Janvier14} showed that currents restricted in localized ribbons are consistent with the overall free energy decrease during a flare, while the decrease of the electric currents at the ribbon hooks during an eruption can be related to the expansion of the flux rope in the corona during the eruption \citep{Barczynski20}. \citet{Savcheva16} and \citet{Janvier16} demonstrated, by a combined analysis of the photospheric traces of an eruptive flare, in a complex topology, the tight link between direct measurements of electric currents, topological magnetic structures and flare emission. Focusing on X-ray emission, \citet{Musset15} showed that the elongated  emissions from both thermal and non-thermal electrons overlay the elongated narrow electric current ribbons observed at the photospheric level. \citet{Dalmasse15} demonstrated that neutralized currents are in general produced only in the absence of magnetic shear at the photospheric polarity inversion and conclude that photospheric flows, such as magnetic flux emergence, can build up net currents in the solar atmosphere. These results, providing support for eruption models based on pre-eruption magnetic fields that possess a net coronal current, are essential to understand the onset of solar eruptions (cf. Sections \ref{sec:solar_eruptions} and \ref{sec:spaceweather_flares}). 

\begin{figure*}[ht!]
    \centering
    \includegraphics[width=\linewidth]{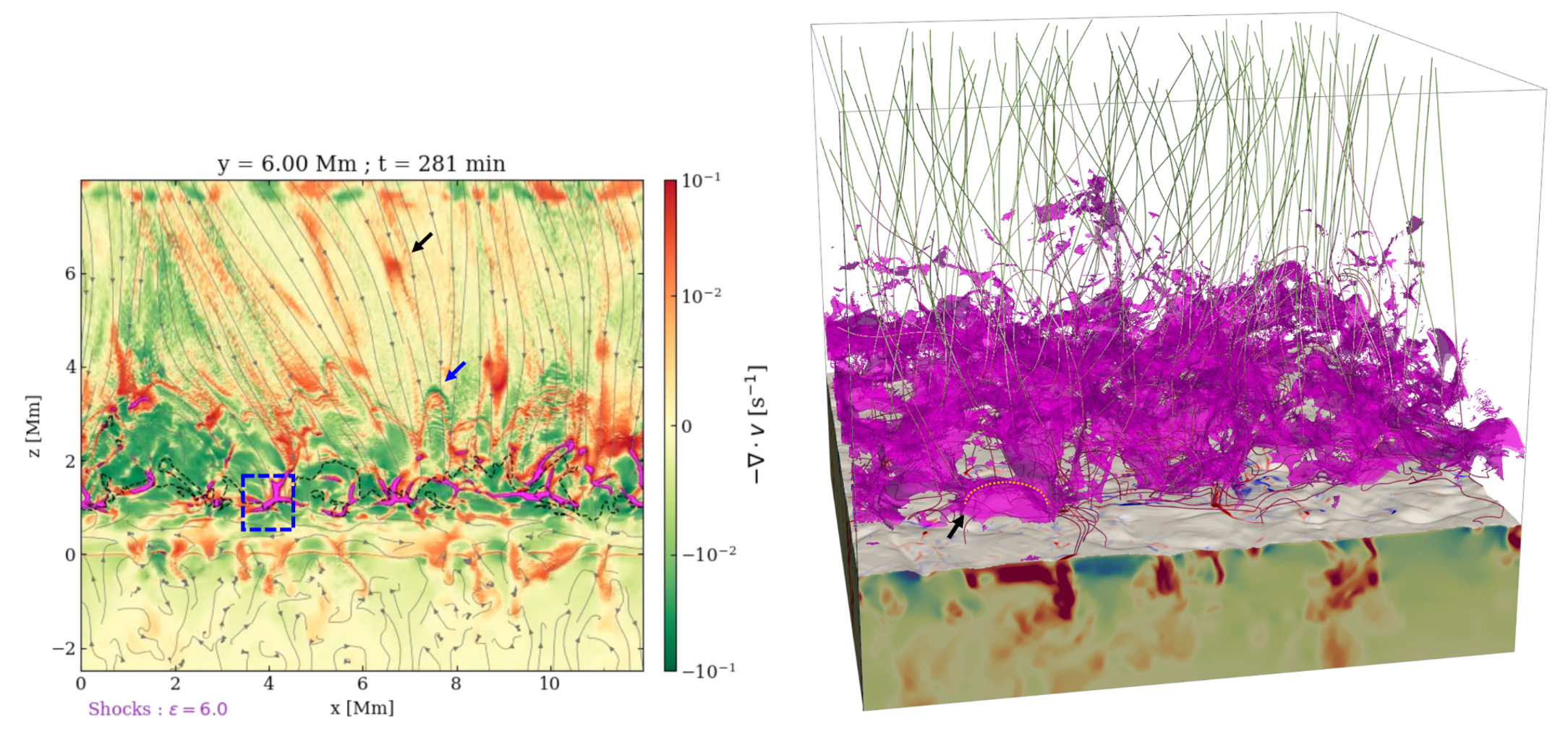}
    \caption{Shock locations in a QS coronal hole numerical \textit{Bifrost} model. The left panels shows regions of compression (red) and expansion (green) in the chromosphere and highlights shock dynamics (purple). Magnetic field lines indicate the different configuration. The dashed line outlines a chromospheric shock region, close to the $\beta=1$ surface, while arrows mark compression sites associated with wave propagation (black) and rising overdensities (blue). The 3D rendering (right) illustrates the distribution of shock fronts and their dome-like geometry. Associated animations with these figures are available in \cite{norazChromosphereQuietSun2026}, from which this figure is adapted.}
    \label{fig:ChromoShocks}
\end{figure*}

Accurate electric current estimations
require combining high polarimetric precision with high spatial
resolution at multi-height measurements simultaneously. This task, which is very challenging, will be a core objective of EST (cf. Section 3.6 of the \citetalias{ESTSRD}). Electric current are indeed derived from the curl of the magnetic field. High spatial resolution is thus necessary in order to have proper horizontal derivatives to estimate correctly the vertical component. The transverse component of the electric current requires estimations of the magnetic field at different layers of the solar atmosphere \citep{Bommier13,Molodij13}. Thus it is mandatory to obtain spectropolarimetric measurements in different lines, hopefully simultaneously, with atomic transitions occurring in distinct physical condition, i.e. at different height of the stratified solar atmosphere (cf. Tables 3.6.1-3.6.3 of the \citetalias{ESTSRD} for examples of relevant observing programs). 
EST's ability to operate all of its instruments simultaneously is a unique feature that makes it very suitable for this task of observing multiple heights and, combined with the high spatial resolution and polarimetric sensitivity, deliver unique datasets with three-dimensional and temporal information of the magnetic field, and the electric currents, crucial to uncover their exact role in the trigger of solar active events.

In addition to the expected discoveries
about the nature of the couplings between the different layers of the magnetised solar atmosphere during flares and eruptions, these novel measurements will also open a new window to better understand and predict solar active phenomena, in the framework of space weather \citep[][see also Section \ref{sec:spaceweather_ambiguity}]{Guennou17}. As discussed in Section \ref{sec:spaceweather_flares}, electric current measurements are also key for the early prediction of solar eruptions.

\subsection{Small-scale dynamics and heating of the chromosphere}\label{sec:solar_chromo}

Convective motions of the solar envelope continuously inject energy into the atmosphere of the Sun. Part of this energy flux is converted into heat locally, as revealed by the rise in temperature starting in the chromosphere. Owing to its relatively high density, this layer requires substantial energy input to maintain its thermal structure, as shown by semi-empirical models \citep[e.g.][]{withbroeMassEnergyFlow1977,vernazzaStructureSolarChromosphere1981}. The chromosphere thus plays a critical role in regulating the mass and energy flux injected into the corona and, ultimately, into the heliosphere.

Energy injection manifests through the propagation of MHD waves, some of which steepen into shocks due to the sharp density and temperature gradients near the photosphere \citep[][]{biermannZurDeutungChromosphrischen1946,schwarzschildNoiseArisingSolar1948}, and through the formation of current sheets induced by magnetic braiding and flux emergence \citep[][]{parkerTopologicalDissipationSmallScale1972,parkerRapidDissipationMagnetic1982}. However, quantifying the respective roles of these processes remains challenging because of the highly dynamic and turbulent nature of the chromosphere \citep[][]{carlssonNewViewSolar2019}, especially when considering the diversity of magnetic configurations over the solar surface and during the magnetic cycle. 

Recent numerical and observational advances have begun to disentangle the relative contributions of waves, shocks, and reconnection to the chromospheric energy budget \citep[e.g.][]{abbasvandChromosphericHeatingAcoustic2020,cherryDecomposingWaveActivity2026,norazChromosphereQuietSun2026}.  
Figure~\ref{fig:ChromoShocks} is adapted from \cite{norazChromosphereQuietSun2026}, where they highlight the dynamics of MHD shocks in current numerical setups of the solar chromosphere. The authors developed physics-based criteria to track and decipher the respective roles of different processes in the dynamics and heating of chromospheric radiative-MHD models. This analysis, among others in the community \citep[e.g.][]{Robinson22,Robinson23}, helps our understanding of chromospheric physics, its coupling to the higher layers, and the subsequent parametrizations of it, given our numerical limitations.

Overall, models remain currently limited by the assumptions of MHD formulations \citep[e.g.][]{martinez-sykoraChromosphericHeatingLocal2023} and the present lack of strong magnetic field strength and topology constraints. By increasing the spatial and temporal resolution, EST will help us refine our constraints on chromospheric dynamics and models, such as MHD waves, shocks and current sheets roles. This is the object of multiple top-level science goals of EST: cf. Sections 2 \& 3 of the \citetalias{ESTSRD}.
Precise measurements of field vectors (see also Section ~\ref{sec:solar_electriccurrents}) will indeed be essential to study how this dynamics evolve over the solar magnetic diversity, and especially how MHD waves are refracted, reflected, and converted between acoustic and magnetic modes in the mid and upper chromosphere \citep[e.g.][]{enerhaugIdentifyingMagnetohydrodynamicWave2025}.

Although the chromosphere is an optically thick medium, the carefully selected set of chromospheric spectral lines observed by the EST instruments will allow us to probe different formation heights and thus retrieve key plasma and wave properties throughout this complex layer (cf. observing programs/tables 2.4.1, 3.1.1, 3.4.1, 3.4.2 and 3.5.1 of the \citetalias{ESTSRD}). By combining high-resolution, multi-line spectropolarimetry, EST will allow us to directly test and refine numerical models, quantify the efficiency of various mechanisms in the chromosphere, and its role in higher layers’ response. In particular, it is known that the chromospheric temperature directly influence the formation height of the transition region, which further dictates the mass loaded into the overlaying corona, and therefore sets its temperature in a highly non-linear manner (e.g. \citealt{gudiksenInitioApproachSolar2005a}; Noraz et al. in prep). Such fundamental physics constraints are still lacking, but yet are crucial for setting realistic condition in coronal models (e.g. for the French ISAM code, see Sections ~\ref{sec:solar_abundances} and \ref{sec:solar_helium_sw}), which are in particular key for reliable solar wind driving in space weather forecasting \citep[e.g.][see also Section \ref{sec:spaceweather_connectivity}]{parentiValidationWaveHeated2022,brchnelovaCOCONUTMFTwofluidIonneutral2023,brchnelovaConstrainingInnerBoundaries2025,wangMHDModellingOpen2026}.

\begin{figure}
    \centering
    \includegraphics[width=.99\linewidth]{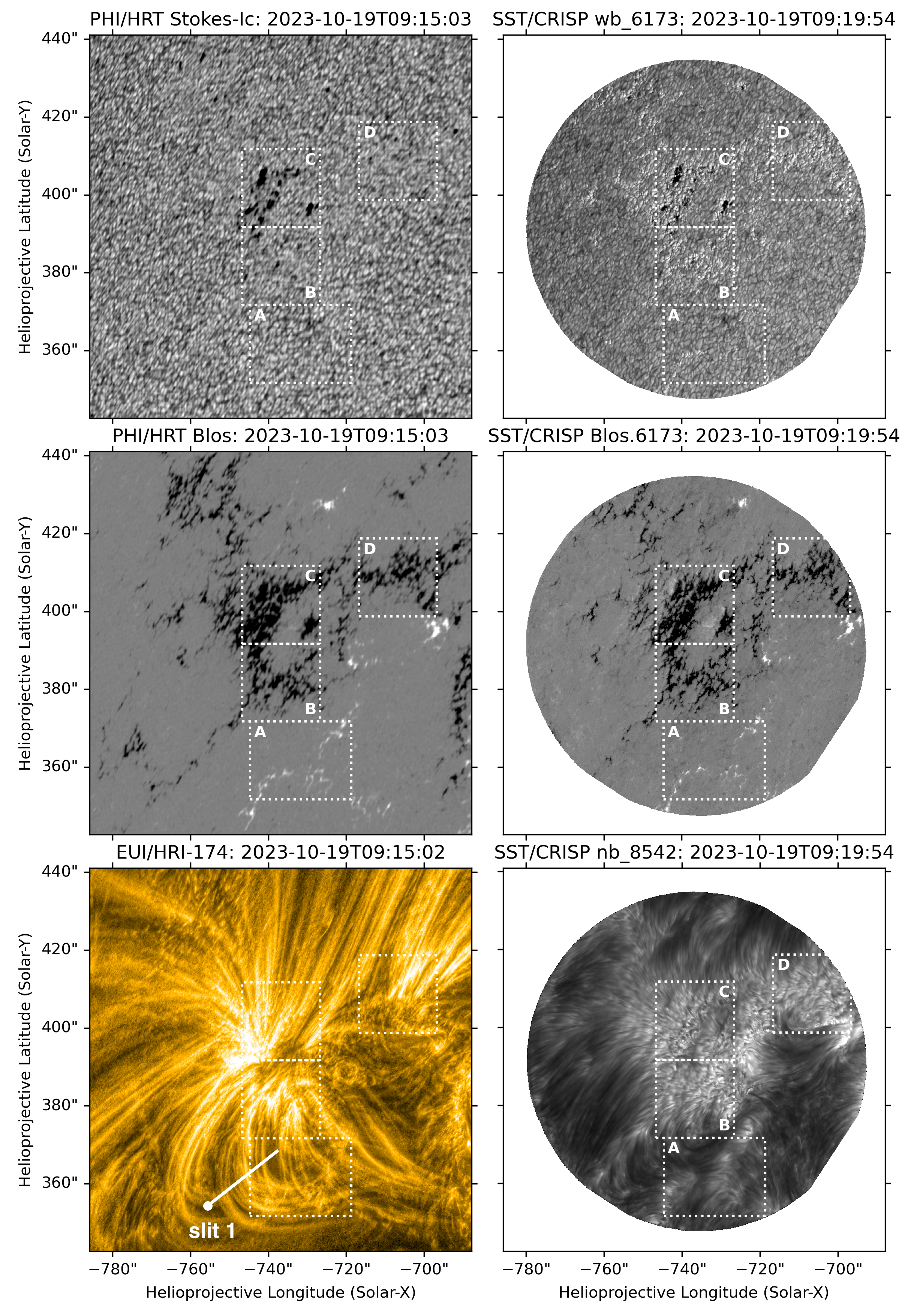}
    \caption{Zoom-in view of active region 13470 observed on October 2023 conjointly by SolO (left column) and SST (right column). The first row shows the photosphere in the continuum emissions of neutral iron, the second row the inferred line-of-sight magnetic field, and the third row the chromosphere in the H$\alpha$ line core (right panel) and hot corona at 174$\AA$. Figure taken from \citet[][]{Poirier2025}.}
    \label{fig:SST-SolO_coordination}
\end{figure}

\subsection{The photosphere-corona connection: excitation of coronal oscillations by photospheric motions} \label{sec:solar_coronaloscillations}

Coronal loops constitute a significant part of active regions. They are large-scale magnetic structures that connect back to the solar surface, and appear as bright arch-like features in extreme UV imagers due to their high temperature and density \citep[see e.g. the review by][]{Reale2014}. Coronal loops are known to be efficient wave guides, by channeling and accumulating wave energy in the corona. Although there is now a consensus that waves are major contributors to coronal heating, we still have a poor understanding of the nature of coronal waves, as well as of their excitation mechanism(s) and driver(s) \citep[see e.g. the reviews by][]{Nakariakov2020,Doorsselaere2020}. There is nonetheless a strong indication that the photosphere is involved with its perpetual motions due to convection and transport of magnetic flux for instance \citep[see e.g. the review by][, and references therein]{Rincon2018}. EST will be crucial in that context by allowing us to study in more detail how this large energy reservoir in the photosphere can effectively serve at exciting coronal waves, and how wave energy is transported throughout the chromosphere.

Coronal waves have first been detected as transverse oscillatory motions of the loop axis in EUV images \citep[][]{Nakariakov1999,Aschwanden1999}. They are essentially standing MHD waves associated with the kink modes of coronal loops \citep[see e.g. the review by][]{Nakariakov2021}. Most of these transverse oscillations do not exhibit decay over their apparent lifetime \citep[see e.g.][]{Anfinogentov2015,Zhong2022}, suggesting that they must be fed by a quasi-continuous source of energy to compensate for their damping. 

In order to study the possible interplay between the photospheric and coronal dynamics, specific observations are needed of these two regions including the various atmospheric layers in between that constitute the lower solar atmosphere. This requires extensive observation campaigns coordinated between various instruments, often from both space and ground to get access to a broad wavelength coverage. Figure ~\ref{fig:SST-SolO_coordination} shows an example of such observations for an active region, a campaign that was led by the Solar Orbiter ESA mission \citep[SolO;][]{Muller2020} and coordinated with multiple spacecrafts and ground-based telescopes including SST.
Since then, regular observation campaigns of this kind are coordinated with a strong involvement from the french community (e.g. LPC2E and IAS), that demonstrates the importance of concerted efforts between space- and ground-based assets.

SST is a highly versatile and relevant telescope for the observation of diverse solar features at high precision, that has already proven to be very effective in coordination with space-based observations \citep[see e.g.][]{DePontieu2014,Antolin2015,Froment2020,Rouppe2020,Poirier2025,Joshi2025}. In the work shown in Figure ~\ref{fig:SST-SolO_coordination}, SST provided crucial high spatial and temporal resolution spectro-polarimetric observations of the photosphere and chromosphere that completed the SolO instruments. The spectro-polarimetric observations from SST were used to track photospheric motions at the footpoints of active region loops in different photospheric configurations such as pores, plages, and enhanced networks, as well as sunspots (not shown here). The same active region loops were then analysed from a coronal perspective, by cutting artificial slits perpendicularly to the loops seen in the extreme ultraviolet (EUV) images obtained by SolO (see bottom left panel of Figure ~\ref{fig:SST-SolO_coordination}) to derive their transverse (kink) oscillation properties. No clear correlation could be found between the properties of the motions at the photosphere and those of the coronal oscillations above. This is indicative of the existence of a complex excitation mechanism of these coronal waves, and reveals that the photosphere-corona connection is very delicate to make in that context.

Among the various complexities encountered are the waves which evolve significantly during their propagation throughout the chromosphere and transition region, where a variety of physical processes occur such as the transfer and coupling between acoustic and magneto-acoustic waves, mode-conversion, steepening, reflection, refraction and damping. The chromosphere is also known to be a very dynamic and turbulent layer by itself, affecting the propagation of these waves. EST can largely contribute to improve our understanding of the photosphere-corona connection in the wave context especially, by providing key measurements on how the energy is transported throughout the chromosphere a major science objective of EST (see e.g. Sections 2 \& 3 of the \citetalias{ESTSRD}) with dedicated observing programs planned (cf. Tables 2.4.1, 3.4.1, 3.4.2, 3.5.1 \& 3.5.2 of the \citetalias{ESTSRD} for an example). The above study also lacked a precise way to trace back the footpoint of the active region loops seen in EUV down to the photosphere, because magnetic field inferences from chromospheric lines such as Ca II H \& K via spectropolarimetry are still not accurate enough due to the limited SNR even for current 1-m class solar telescopes. In that regards EST will be highly valuable if not critical to improve our understanding of the photosphere-corona connection, by providing unprecedented high-sensitive measurements of these chromospheric lines allowing for more precise magnetic field tracing in the chromosphere (like in active region plages, see e.g. observation program/Table 3.8.1 of the \citetalias{ESTSRD}).

\begin{figure}[ht!] 
	\resizebox{\hsize}{!}
	{\begin{tabular}{c} 
		\centerline{\includegraphics[scale=0.4, trim=0 0 150 0,clip]{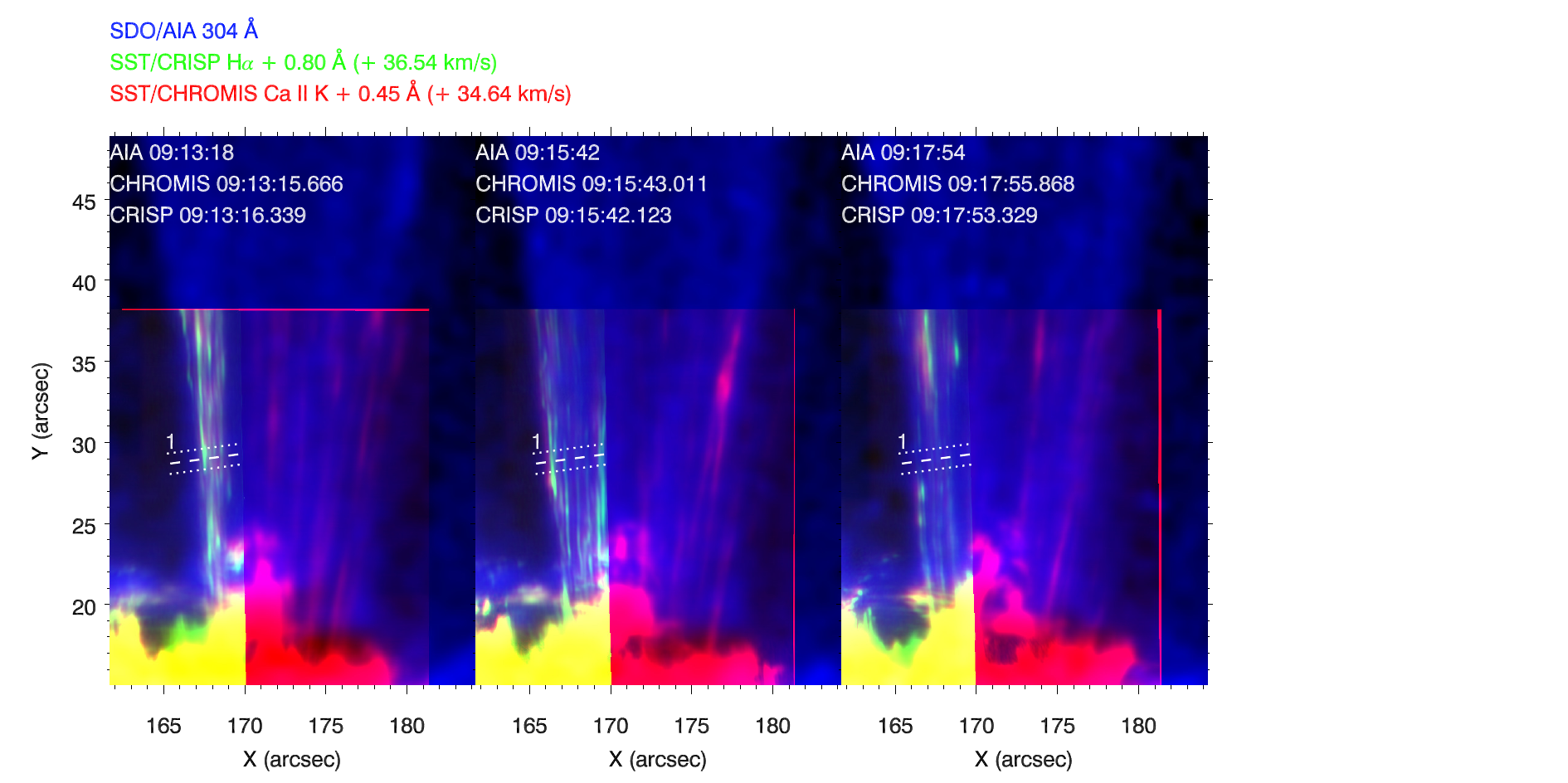}} \\
		\centerline{\includegraphics[scale=0.3, trim=0 0 0 150,clip]{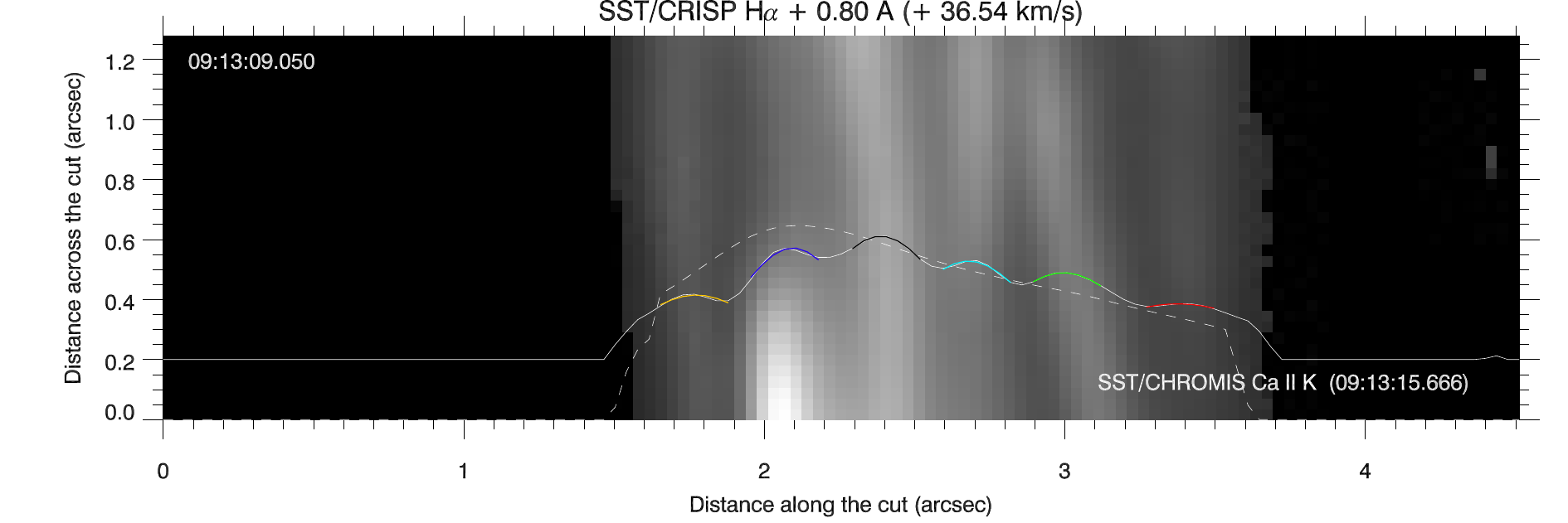}}
	\end{tabular}
	}
\caption{Fine-scale coronal rain observed near the footpoint of large coronal loops with a ground-based telescope. Top panel: RGB image combining one spectral band of SDO/AIA,  $\rm{H\alpha}$ observations from SST/CRISP, and Ca II K observations from the SST/CHROMIS. Bottom panel: $\rm{H\alpha}$ slice using the path (1) demarcated by dotted lines in the top panel. Six rain strands are clearly visible with widths of about 0.2 arcseconds (i.e. 140 km). Figure adapted from \citet[][]{froment_multi-scale_2020}. }
\label{fig:coronal_rain}
\end{figure}

\subsection{Multi-phase plasma in the solar atmosphere: coronal rain and its relationship with coronal heating mechanisms} \label{sec:solar_coronalrain}

Coronal rain is a frequent feature of active regions. Even though it is best observed in emission lines typical of the transition region and chromosphere lower temperatures, coronal rain is indeed produced at coronal heights. A local thermal instability drives the formation of cool ($\sim 10^4$~K) and dense ($\sim 10^{-11}$ cm s$^{-3}$) condensations that are then about a 100 times cooler and denser that the surrounding coronal material \citep[see for example the review by][]{antolin_multi-scale_2022}. Coronal rain shares similarities with other types of multiphase plasmas that are ubiquitous across the Universe -- precipitation-regulated feedback cycles that drive star and galaxy formation, solar and stellar prominences \citep[see for example the white paper by][]{antolin_cool_2022}. 
Coronal rain appears on the timescale of minutes to hours, in the form of partially ionised \citep[e.g.,][]{braileanu_coronal_2025} clusters, or "showers", of small-scale condensations \citep[few hundreds of km, e.g.][and Figure~\ref{fig:coronal_rain}]{froment_multi-scale_2020}. 
It is best observed in emission while the coronal loops appear above the limb, while it is more challenging to observe coronal rain as small-scale absorption features on the disk \citep[e.g.,][]{kriginsky_thermal_2024}.
Counter-intuitively, these condensations appear under specific atmospheric heating conditions. The thermal instability, also called "thermal runaway" or "catastrophic cooling" driving the formation of the condensations, is triggered during thermodynamic cycles that are called thermal non-equilibrium \citep[TNE, e.g.,][]{antiochos_model_1991}. Coronal loops can undergo TNE cycles under quasi steady (i.e.,  high frequency compared to the cooling timescale) and stratified (most the heating is concentrated at the footpoint of the structure) heating conditions. Such a heating will drive evaporation of the chromospheric plasma, filling the loops with dense and hot material. This material, now at coronal heights, will cool down because of the insufficient volumetric heating in this region. This cooling will further proceed as radiative losses will increases as the density increases and the temperature decrease. This will trigger the formation of the condensations.
Coronal rain is also observed during the late phase of solar eruptions \citep[e.g.,][]{foukal_magnetic_1978,schmieder_relation_1995, jing_unprecedented_2016, mason_rain_2022}. However, its formation mechanism is still under debate, as the impulsive nature of the energy deposition driving the chromospheric evaporation as to be reconciled with the TNE model \citep{reep_electron_2020, benavitz_spatiotemporal_2025}. 

\begin{figure*}[ht!]
    \centering
    \includegraphics[width=0.95\linewidth]{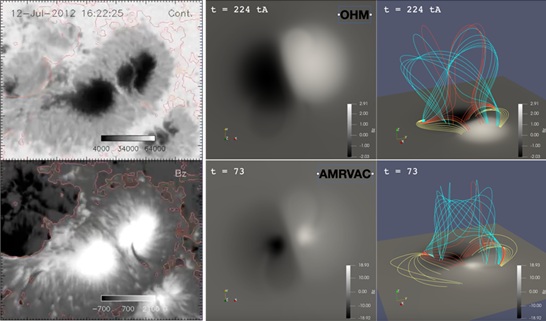}
    \caption{Two simulations using the OHM and AMRVAC codes of eruptive twisted flux tubes (right panels), and a comparison of the synthetic magnetograms (middle panels) with an SDO/HMI magnetogram (left) showing the “scar” of the sunspot (in dark, on the positive-polarity spots in white), identified through model–observation coupling as being located at the footpoints of the eruptive flux tubes \citep[][]{Xing24}.}
    \label{fig:magneticscar}
\end{figure*}

Coronal rain was first discovered with $\rm{H\alpha}$ ground-based observations by \citet{kawaguchi_observed_1970} and \citet{leroy_emissions_1972}. Modern high-resolution instruments now allow the community to use these small-scale features to study multi-stranded nature of coronal loop \citep{antolin_multithermal_2015}. The braiding of coronal loops strands is a key component of the nanoflare heating model \citep{parker_nanoflares_1988}. Recent observations from the 1.6-m GST, thanks to their new coronal adaptive optics capabilities, resolve coronal rain blobs down to the diffraction limit of the telescope \citep[87 milliarcseconds, i.e. 64 km, in $\rm{H\alpha}$,][]{schmidt_observations_2025}. 
In numerical simulations, rain blobs have also been shown to have widths below 100 km \citep[e.g.,][]{fang_multidimensional_2013,li_coronal_2022}. With higher-resolution, coronal rain blobs reported gets smaller, it thus suggest that we are limited by the spatial resolution. 
Coronal rain observations can also be used to determine quantities that are otherwise very difficult to measure, such as the coronal magnetic field.
Spectrolarimetric signals of coronal rain observations have been used to map the LOS component of the magnetic field (weak-field approximation) \citep{schad_vector_2016,kuridze_mapping_2019,kriginsky_magnetic_2021}. This method requires a high signal-to-noise ratio.

The French community has been highly involved in the study of the coronal manifestation of TNE that are long-period EUV pulsations. These pulsations are reflecting the temperature and density evolution in the corona \citep[e.g.][]{auchere_long-period_2014, froment_evidence_2015, froment_long-period_2017,pelouze_spectroscopic_2020}. Our investigation allowed us to link long-period EUV pulsations to coronal rain, providing the observational evidence that TNE and coronal rain are indeed two faces of the same coin \citep{auchere_coronal_2018,froment_multi-scale_2020, sahin_spatial_2023}. These advances were part of the work of recent international ISSI teams led by P. Antolin and C. Froment\footnote{\url{https://www.issibern.ch/teams/observecoronloop/}} \footnote{\url{https://teams.issibern.ch/multiphaseplasmas/}}. We have experience on coordinating ground-based and space-based coronal rain observations, proven to be complementary to study these multi-thermal phenomena \citep{froment_multi-scale_2020}.
We thus envision future coordinated studies between EST and space-borne instruments to further connect this large range of thermal and spatial scales, in particular for coronal heating studies. For example, estimating the presence of braiding at sub-arcsecond scales is essential in assessing the coronal heating mechanisms such as braiding-induced magnetic reconnection. Recent studies also estimate the energy release caused by the impact of coronal rain in the chromosphere \citep{wachira_compression_2025} which demonstrate how coronal rain properties can be used to infer coronal heating parameters. 
Even thought EST will not be tailored to carry out coronal observations, coronal rain observations are commonly performed using chromospheric lines, slightly off the disk \citep[e.g.][]{froment_multi-scale_2020} or on disk \citep[e.g.][]{antolin_-disk_2012}, which is compatible with EST programs using imaging and spectroscopic modes in the $\rm{H\alpha}$ and Ca II H lines (cf. Tables 4.4.1, 4.5.1, 4.6.1, 4.7.1, 7.1.2, 7.4.1, 7.5.1 of the \citetalias{ESTSRD} for examples of relevant observing programs).

\begin{figure*}[ht!]
    \centering
    \includegraphics[width=\linewidth]{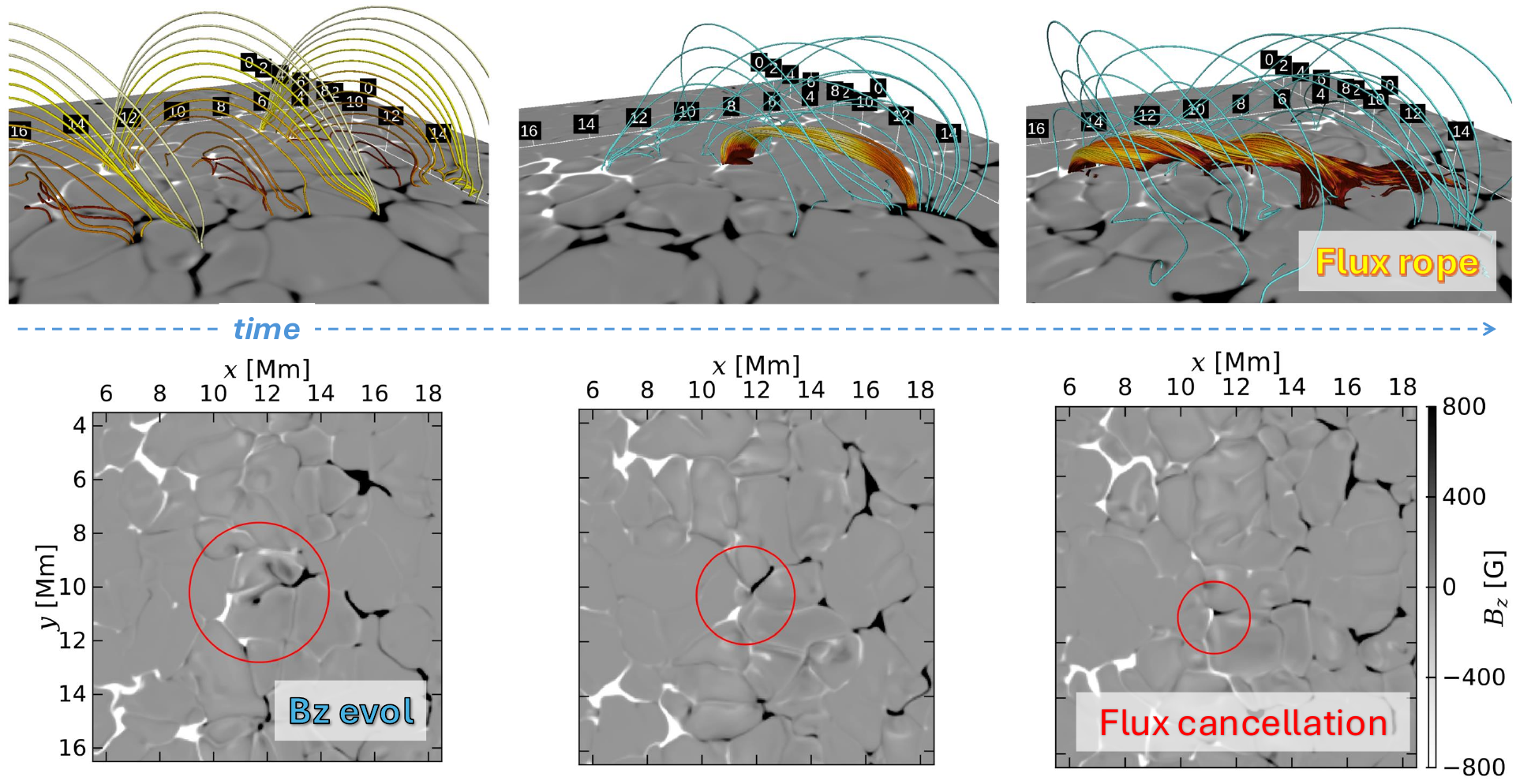}
    \caption{
     Flux rope formation via sparse flux-cancellation in a \textit{Bifrost} simulation. Top row show a time sequence of 3D rendering illustrating how a initially un-twisted magnetic field (top-left panel) can self-consistently turn into a flux-rope via granulation forcing. The bottom row illustrate photospheric magnetic maps of the vertical component $B_z$ during this time sequence and highlights flux cancellations during the reconfiguration of the magnetic topology \citep[][]{furusethFluxRopeFormation2025}. }
    \label{fig:fluxropeformationby}
\end{figure*}

\subsection{Low atmosphere signatures of coronal eruptive structures and phenomena} 
\label{sec:solar_eruptions}

Solar eruptions represent the most violent and dramatic forcing on Earth’s magnetic environment within the context of space weather (see Section \ref{sec:spaceweather}). Solar eruptions consist, among other elements, of a flare — a sudden increase in electromagnetic emission across the entire spectrum — and of an ejecta, a magnetic structure carrying coronal plasma into interplanetary space. The former is the result of resistive instabilities associated with magnetic reconnection. Depending on the type of event, the latter results either from a loss of equilibrium \citep[e.g. for coronal mass ejection][]{Aulanier10,Aulanier14} or from plasma acceleration induced by an Alfvénic wave \citep[e.g. for jets][]{Pariat09,Touresse24}. For all these phenomena, the respective roles of resistive and ideal effects are still poorly understood, and the causal relationships between these two types of effects remain to be established. Understanding the three-dimensional dynamics of the coronal magnetic field  before and during eruptions is fundamental \citep{Dudik25}. Three dimensional effects have a physical (and not only morphological) impact on the trigger and properties of the eruption : they determine the 3D regime and the global role of magnetic reconnection involved in eruptive events, and thus must be included in the standard flare model \citep[][]{Aulanier12,Aulanier13,Janvier13,Janvier15}. Photospheric and chromospheric observations provides a wealth of information on such 3D properties and the dynamics of eruptive structures, fueling numerous research axes in which French researchers have been heavily involved.
    
Moving flare ribbons is one of the most distinctive features of flares, easily observed in most of the chromospheric lines that EST will observe (e.g. observing programs of tables 6.1.1 \& 6.2.1 of the \citetalias{ESTSRD}). The shape, spatial distribution and dynamics of ribbons has been heavily used as a marker of the 3D reconnection processes during eruptions \citep[e.g.][]{Savcheva12,Savcheva15,Savcheva16,Janvier16,Thoen25,Dudik25}. In particular ribbon hooks, located at the tip of some eruptive ribbons are a key marker of the existence of twisted flux rope during the eruptions \citep{Zhao16,Dudik19,Joshi24}. In simulations, \citet{Aulanier19} predicts continuous deformations and a drifting of interplanetary CME flux-rope footpoints whose areas are surrounded by equally evolving hooked-shaped flare-ribbons. From observed data \citep{Barczynski20} showed a decrease of the electric currents in the area surrounded by the ribbon hooks during and after the eruption, interpreted as due to the expansion of the flux rope. 

The very dynamics of the emission within the ribbons is also providing key input on the reconnection mechanism involved during the eruptive phenomena. Patches of emission within flare ribbons are frequently observed to present "motions" within flare ribbons \citep[e.g.][]{Dudik14,Dudik25,Lorincik19}. These displacement corresponds a specific 3D mode of magnetic reconnection, which develops within particular topological structures of the magnetic field: quasi-separatrix layers \citep{Demoulin96,Pariat12,Dudik25}.   
\citet{Lorincik25} recently reported the very first observation of the super-Alfvénic apparent displacements of such flare kernels. These displacements, predicted by \citet{Aulanier06,Pariat06}, corresponds to an intermediate mode of reconnection bridging the gap between QSL reconnection and traditional 2D cut-and-paste reconnection modes.  With its combination of sub-arcsecond spatial resolution, $<10$~s temporal cadence, and simultaneous multi-line spectro-polarimetry in key chromospheric diagnostics (e.g. H$_\beta$, Ca II, He I), EST will allow to track such fast, kernel-size motions in space, height and time (see section 6 of the \citetalias{ESTSRD}). By helping to resolve the temporal evolution of reconnection modes at the spatial and temporal scales predicted by 3D models, this will enable quantitative comparisons with the super-Alfvénic kernel motions reported by \citet{Lorincik25} and with the drifting QSL footprints anticipated in numerical simulations \citep[][]{Aulanier06,Aulanier19}.

 Photospheric magnetic field maps in active region provides information on pre-eruptive structures. \citet{Joshi21} found evidence in magnetograms, of the pattern of a long sigmoidal flux rope along the polarity inversion line between two emerging flux region, flux rope that eventually induced a jet. \citet{Xing24} focuses on arc-shaped structure intruding in sunspot umbras, so-called sunspot scars. Sunspot scars displays a more inclined magnetic field with a weaker vertical component and a stronger horizontal component relative to that in the surrounding umbra and is manifested as a light bridge in the white light passband. Sunspot scars are located close to the (pre-)eruptive flux rope footpoint. They provide a new perspective for the identification of pre-eruptive and CME flux rope footpoints, as well as new methods for studying the properties and evolution of pre-eruptive structures and CMEs with photospheric observations only. Observation of light bridges is particularly relevant for ground based observation such as the one provided by EST (cf. Section 4.6 and Table 4.6.1 of the \citetalias{ESTSRD}). EST is designed to excel at studying these intermediate-spatial-scales structures.
  
  Colliding sunspots and sparse magnetic cancellations along polarity inversion lines, observed at the photospheric level, within and between active regions are essential markers of the formation of pre-eruptive structures (sheared arcade and/or flux ropes). Using the Bifrost code, \citet{furusethFluxRopeFormation2025} have carried numerical simulation of the self-consistent formation of a twisted magnetic flux rope within the framework of the flux-cancellation model, while including a natural magnetoconvective forcing, illustrated in Figure~\ref{fig:fluxropeformationby}. This brings the theory of \citet{vanBallegooijen89} back to light, supporting the possibility flux rope formation via sparse flux-cancellation resulting from self-consistent solar surface convective driving. Study of flux cancellation events in the EST high resolution magnetic field maps (cf. Section 1.5 \& Table 1.5.1 of the \citetalias{ESTSRD}) shall provide numerous example of these flux rope formation processes and better understand the formation of magnetic structures susceptible to trigger space-weather-effect-inducing eruptions (see Section \ref{sec:spaceweather_flares}). 
    
\begin{figure}[ht!]
    \centering
    \includegraphics[width=\linewidth]{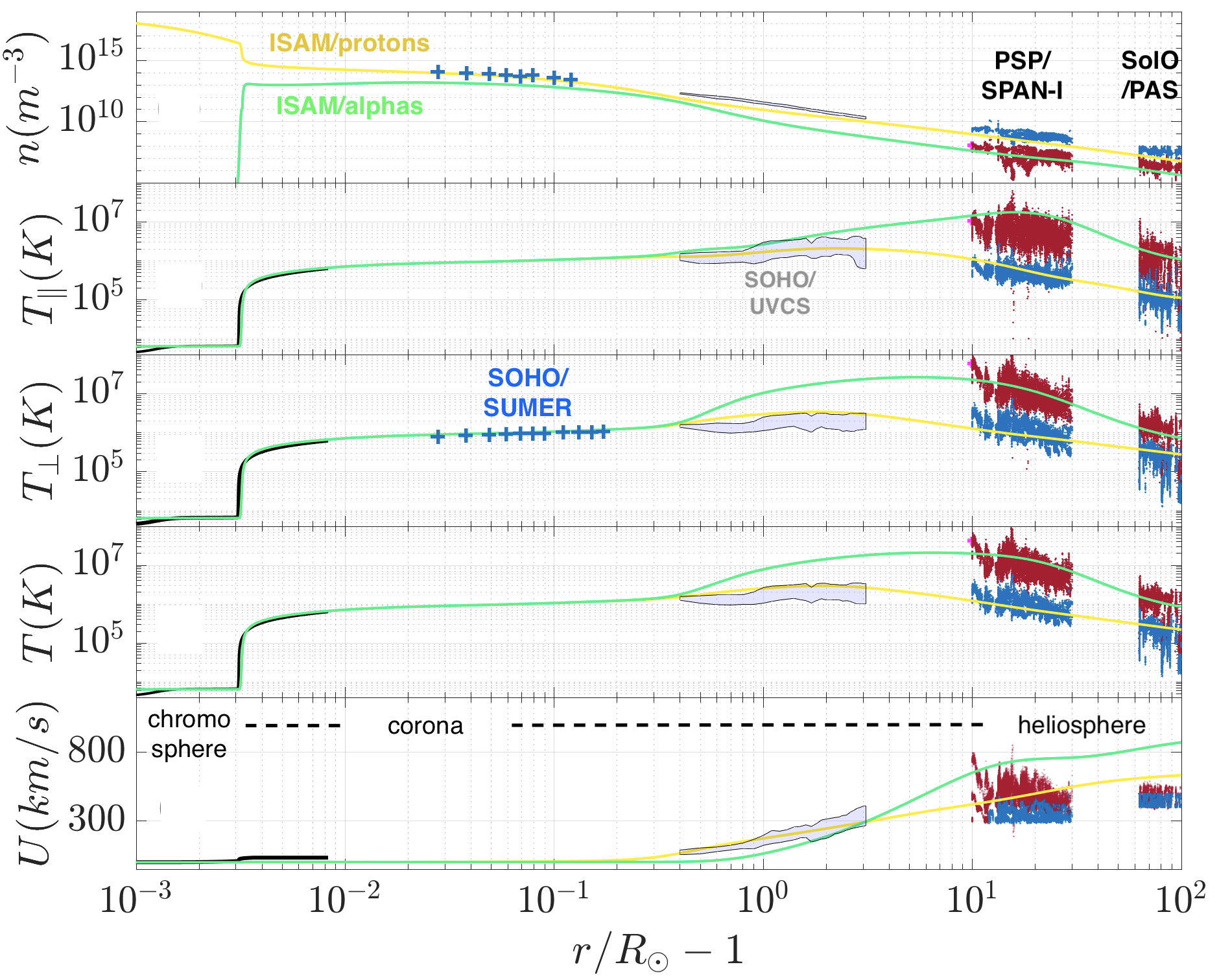}
    \caption{Slow solar wind ISAM 1-D simulations including protons and alpha particles, compared to remote-sensing and in-situ measurements \citep[see][for the latest results]{Lomazzi2026}. PSP and SolO in-situ measurements are given for both protons (blue dots) and alpha particles (red dots). The various panels show respectively, the number density, the parallel and perpendicular component of the temperature (with respect to the magnetic field), the total temperature, and the bulk velocity.}
    \label{fig:ISAM_He_sw}
\end{figure}

\subsection{Modulation of the helium abundance in the solar wind} \label{sec:solar_helium_sw}

If global coronal and solar wind models now fairly reproduce the overall bulk properties of the solar wind. However, most of these models do not include helium yet, whereas it can make up to $20\%$ of the total mass budget and must therefore be accounted for. The abundance of helium has been seen to vary significantly from one solar wind to another, going from about $1\%$ in the slow winds \citep[][]{Kasper2007,Sanchez-Diaz2016} to about $4-7\%$ in the fast winds \citep[see e.g.][]{Stansby2020}. A modulation of helium has also been observed consistently throughout the solar cycle, especially for slow winds that show a strong correlation of the helium abundance with the number of sunspots visible on the surface \citep[][]{Kasper2007,Alterman2025}. These observations tend to favor a source-dependent theory of the slow solar wind \citep[e.g.][]{Baker2023,Yardley2024}, where the solar wind properties (including its helium abundance) are largely determined by the properties of the source region in the corona and below in the chromosphere. EST will be ideally suited to provide new observational constraints of the chromosphere, supporting state-of-the-art modeling efforts as well 
(cf. top level science objective 5.3 of the \citetalias{ESTSRD}).

The modelling of helium in the solar wind is highly challenging, because of the multi-scale nature of the problem and because helium must be solved self-consistently together with the main hydrogen and electron plasma, all the way from the source region up to the heliosphere where it is measured. This is nonetheless possible in highly tailored numerical approaches, such as the high-order 1-D hydrodynamics code called ISAM (for IRAP Solar Atmospheric Model) developped in France at IRAP over the last six years and led to three Ph.D. thesis \citep[][]{Lavarra2022THESIS,Poirier2022THESIS,Lomazzi2025THESIS} and a series of papers \citep[][]{Lomazzi2025,Lomazzi2026,Poirier2026}. An example of ISAM solar wind simulations including alpha particles (doubly ionised helium) is shown in Figure~\ref{fig:ISAM_He_sw}. This work shows that the simulated bulk properties of both protons (singly ionised Hydrogen) and alphas particles can fairly well reproduce observations, including the recent in situ measurements obtained by the SolO and Parker Solar Probe (PSP) missions with instruments that are heavily supported by the French scientific community.

This required, however, a fine tuning of how the plasma is heated, especially the location and distribution over the different species. We have concluded that the way the energy sources and sinks are set lower down in the chromosphere and transition region can have a dramatic effect on the terminal properties of the simulated solar wind protons and alphas. Observations of the solar chromosphere are scarce, and its opacity to radiation makes any diagnostic of the plasma properties difficult. In particular, the ISAM simulations would greatly benefit from having better observational constraints on how waves are generated, propagate and eventually heat up the plasma in the chromosphere. This would require high-cadence measurements at different heights simultaneously, which EST will be able to provide thanks to its distinct channels covering from the solar photosphere up to the upper chromosphere (cf. Tables 2.4.1, 3.4.1 \& 3.4.2 of the \citetalias{ESTSRD} for example of relevant observing programs). Wave-based models are also highly dependent on the magnetic field properties, which remain poorly constrained in the chromosphere due to the lack of sensitivity of current ground-based infrastructures to allow accurate magnetic field inferences in that layer. Therefore, EST, with its highly sensitive polarimetric measurements in the chromosphere (see e.g. Table 3.8.1 of the \citetalias{ESTSRD}), will be a highly valuable asset to better constrain state-of-the-art solar wind models developed in the international community and in France, where a better description of the lower solar atmosphere, at the source of the solar wind, becomes essential.

\begin{figure*}[ht!]
    \centering
    \includegraphics[width=0.45\linewidth]{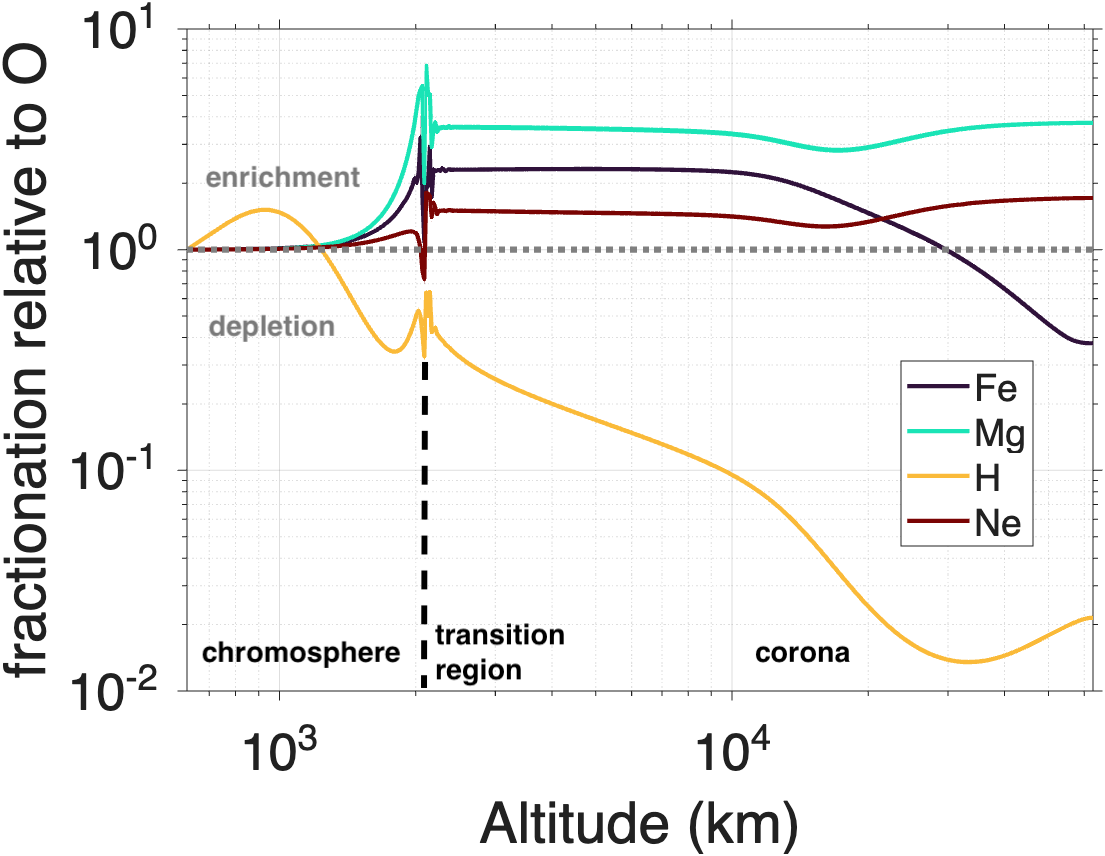}
    \includegraphics[width=0.41\linewidth]{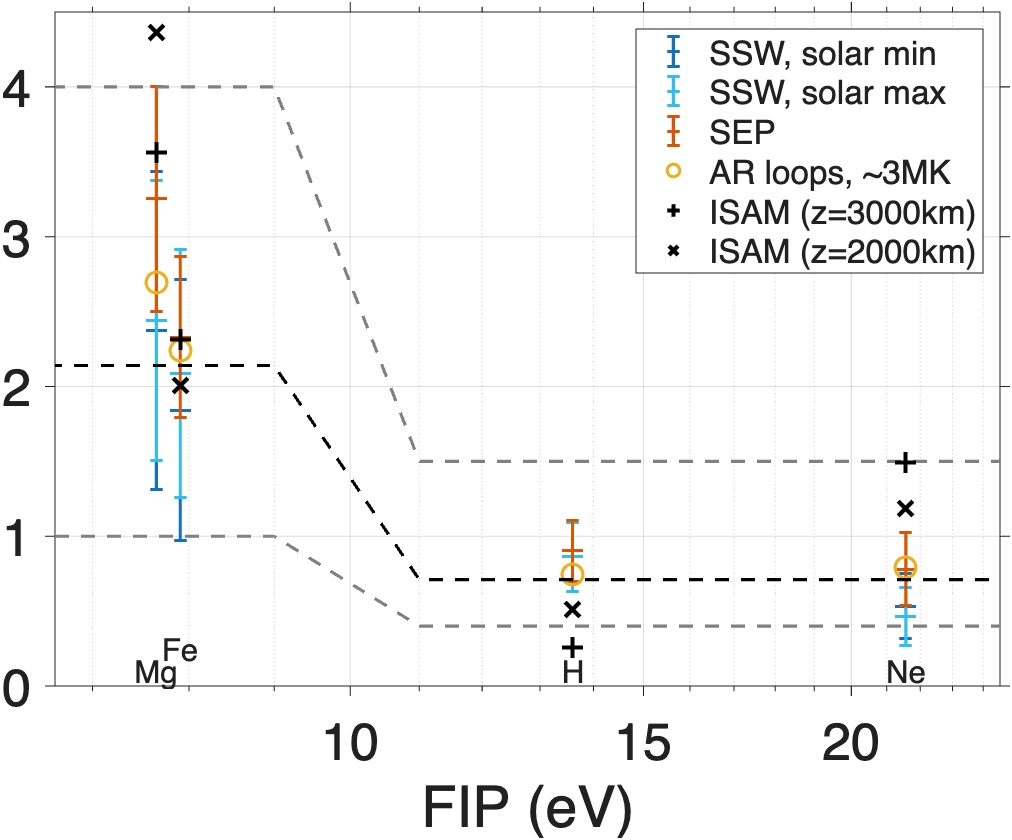}
    \caption{Fractionation of heavy ions in an active region loop modelled with the ISAM code. The plots show the deviation of atmospheric abundances from the photospheric (surface) values, where abundances are here taken relative to Oxygen. Left panel: Evolution of the fractionation with altitude. Right panel: Fractionation as a function of FIP, extracted at two heights in the ISAM simulations at the top of the chromosphere ($z=2000$ km) and the base of the low corona ($z=3000$ km). ISAM results are comparable against observations of slow solar winds \citep[SSW;][]{Steiger2000}, solar energetic particles \citep[SEP;][]{Reames1995}, and active region (AR) loops \citep[][]{DelZanna2013}. Iron and magnesium get already enriched in the chromosphere, reaching coronal values comparable with observations. To be noted that there are large variations within the measurements of fractionation given in the literature, where Iron ions have been found to be enriched up to a factor of 10 in long-lived coronal loops for instance, and where the measured fractionations can get sensitive to height due to gravitational settling overall.} 
    \label{fig:ISAM_FIP}
\end{figure*}

EST should then benefit not only for fundamental research, but also for solar wind research and numerical models that constitute the base of space weather applications. In particular, the modeling of alpha particles can significantly help us to better estimate the terminal solar wind mass flow and hence the dynamic pressure exerted on Earth’s magnetosphere (see \sect{spaceweather_connectivity}). Additionally, the assumed solar wind density and bulk velocity has a great influence on how coronal mass ejections propagate and evolve throughout the heliosphere before they eventually reach Earth (via e.g. erosion and drag as discussed in \sect{spaceweather_connectivity}). The exploitation of novel numerical models like ISAM, that extends all the way from the near solar surface up to the heliosphere, makes it then possible to extend such lower solar atmospheric observations proposed by EST to a broader purpose and scientific context.

\subsection{Transfer of heavy ions into the solar corona and wind} \label{sec:solar_abundances}

In addition to Helium, the solar wind is also constituted of a plethora of heavier ions. Although they only represent a minor contribution to the total number density ($<1\%$), they act as tracers of the source regions and formation mechanisms of the solar wind. Indeed, the composition of the solar wind in heavy ions shows intriguing properties that seem to arise very early on in the solar atmosphere, in the chromosphere especially where the plasma starts to get partially ionised. In that context, EST can provide novel plasma and magnetic field diagnostics that will be critical to better understand the transport of heavy ions in the chromosphere and beyond.

Indeed we tend to see a fractionation of chemical elements based on their first ionisation potential (FIP), a phenomenon referred to as the FIP effect \citep[see e.g.][for a review on the FIP effect]{Henoux1998,Laming2015}. For instance, elements having a FIP lower than $\approx 10eV$ such as iron, magnesium, and silicon,  tend to be enriched compared to elements with a higher FIP such as hydrogen and helium. Coronal spectroscopic observations have revealed that this effect is most pronounced in active regions, where the plasma is confined in large magnetic loops and has time to fractionate \citep[][]{Widing2001,Feldman2003,Baker2013,DelZanna2014,Brooks2015,Mihailescu2022}. Interestingly a similar strong enrichment in low-FIP heavy ions has also been measured in situ far away from the Sun, mostly in slow solar winds \citep[see e.g.][]{Geiss1995,Steiger2000,Stansby2020a}. 

These findings unveil the existence of release/extraction processes such as interchange magnetic reconnection at open/close boundaries to regulate the plasma populations fueling the solar wind \citep{Lynch2023,Koukras2025}. The detection of persistent up-flows along with a strong FIP effect at the boundaries of active regions, where slow solar winds take root, has further shown the importance of active regions in driving slow solar winds \citep{Brooks2011}. Such up-flows shall also be the target of EST which will additionally provide simultaneous magnetic field measurements at the root of active regions (cf. Table 5.3.1 of the \citetalias{ESTSRD}). Therefore, the study of heavy ions proves extremely useful, if not necessary, to efficiently track the solar wind sources \citep{Brooks2015,Yardley2024}. However, we still have a very poor understanding on the dynamic contribution of interchange reconnection to the solar wind, a topic currently under study in several french teams such as LPC2E and LPP. The study of heavy ions can then shed a new light on the interplay between active regions and the solar wind, given that we can understand how these heavy ions get enriched in active regions in the first place.

Although some theories have received significant support recently, there is still no clear consensus or quantitative measurements on the mechanisms that regulate the transport of heavy ions throughout the solar atmosphere. The main difficulty arises from the fact that this fractionation process starts to occur in the partially ionised chromosphere, a very challenging layer to model for which we lack observational constraints. The ISAM model that is under development at IRAP (introduced in Section~\ref{sec:solar_helium_sw}), has been specifically designed to tackle this challenge, by modeling the transport of heavy neutrals and ions in an extended multi-fluid approach. A first test in active region loops gave very promising results on the heavy ion fractionation process \citep[][]{Poirier2022THESIS,Poirier2026}, supporting past studies on the importance of diffusion/collisional processes in the fractionation of heavy ions \citep{Marsch1995,Peter1998b,Killie2007}. 

However, these models, including ISAM, still suffer from a poor treatment of the chromosphere where the fractionation occurs. Semi-empirical profiles, inverted from spectroscopic observations, do exist already and are exploited to constrain the simulations, but they only give a rough approximation of the averaged plasma properties in the chromosphere \citep[see e.g.][]{Vernazza1981,Avrett2008}. More constraints are needed in a dynamic context, and for a variety of different photospheric configurations (quiet-sun, enhanced-network, plage, pores, sunspots). In particular, more observational diagnostics are needed for the processes suspected to influence the heavy ions in the chromosphere either through plasma heating or direct driving. EST can greatly fill this gap and support numerical modelling by providing key measurements of waves (cf. Sections 2.4 \& 3.4 in the \citetalias{ESTSRD}), reconnection events at small-scale (cf. Section 3.5 of the \citetalias{ESTSRD}), and of the temperature and magnetic field properties in the chromosphere (cf. Sections 3.7 \& 3.8 of the \citetalias{ESTSRD}).

Although EST will focus on relatively small field-of-view targets, carrying systematic observations should allow the scientific community to map and build empirical constraints of the above chromospheric mechanisms for the different types of photospheric configurations at the root of active regions. Such an empirical work could then greatly benefit current state-of-the-art models developed within the solar physics community and in France, such as ISAM where the chromosphere must be parametrized. Since heavy ion simulations like ISAM are highly sensitive to the underneath photospheric and chromospheric layers, especially for the wave and magnetic field properties, such simulations can be used to connect EST observations with coronal measurements from other plateforms. For example, extensive heavy ion abundances shall be provided by the slit-spectrometer on board Solar-C \citep[EUVST;][]{Shimizu2021} planned for launch in 2030, a JAXA led mission for which french institutions and science institutes are contributing significantly (financial support from CNES, and instrumental co-leading by the Institut d’Astrophysique Spatiale, IAS).

All of these future prospects should allow us to significantly improve our understanding of the heavy ion composition in the solar atmosphere, benefiting space weather applications as well by allowing a more precise tracking of the solar wind sources (for now mostly based on the magnetic connectivity, see section \ref{sec:spaceweather_connectivity}).


\section{French Space Weather research with EST} \label{sec:spaceweather}

Space weather refers to the characterization of the impact of solar activity on the Earth spatial environment, so that it can be anticipated. Indeed, both continuous interactions through irradiance and the solar wind, and energetic impulsive events such as solar flares and Coronal Mass Ejections (CMEs) can have a strong impact on our technological society, requiring surveillance and alerts. Even though EST is not designed as a space-weather-oriented operational telescope, it can help greatly the research surrounding space weather which is crucial to the improvements of our understanding of these phenomena, and can result in better modelling of the Sun-Earth chain, and thus better alerts for all end-users and better shielding for our installations. 
In the following section, we will discuss in particular the following topics of interest for the french community: the improvement on constraints on the solar wind and solar connectivity, the characterization of recurring active regions, the high-resolution observations of filaments as CME precursors, and finally the improvement of eruption prediction thanks to both better interpretation of the measures of the magnetic field and better modelling of solar flares.


\begin{figure}[ht!]
    \centering
    \includegraphics[width=0.99\linewidth]{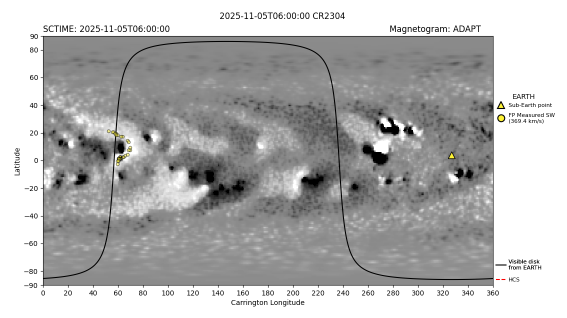}
    \caption{Example of predicted magnetic connectivity from the Earth to the Sun by the \href{https://connect-tool.irap.omp.eu/}{connectivity tool} for November 5 2025 based on a low-resolution magnetogram (GONG-ADAPT). The position of the Earth projected in Carrington coordinates is shown with a yellow triangle. The dark line indicates the Earth field of view. The estimated connectivity on the solar surface is shown with yellow circles. The uncertainties come from the potential field extrapolation of the magnetic field which has free parameters.}
    \label{fig:connectivity}
\end{figure}

\subsection{Solar wind constraints as a link between the solar surface and Earth}
\label{sec:spaceweather_connectivity}

The solar wind is a key physical phenomenon that connects the Earth to the Sun. It is a continuous flow of charged particles (electrons, protons, etc.) ejected from the outer layers of the solar atmosphere. Thus, it acts as a permanent interaction between our planet and our star, which can in itself trigger space weather alerts (high-speed streams can be responsible for geomagnetic storms), but also acts as a key ingredient to understand the propagation of impulsive events towards our planet (especially for CMEs). Several open questions remain regarding the solar wind: how is it accelerated? how to characterize it? what are the regions responsible for the origin of the solar wind? All these questions actually find their answers in the lower atmosphere of the Sun, especially the photosphere and the chromosphere, providing a unique opportunity for EST to contribute to it. With the historical interest of the french community towards understanding the solar wind, with recent observations
\citep[e.g.,][]{Dakeyo2022, Lomazzi2025} and models \citep{Pinto2017, Reville2020, Reville2023, Pellegrin-Frachon2023, Houeibib2025,Baratashvili25}, EST has the potential to nourish many interesting results. We will especially focus on the problematic of the driving and heating of the solar wind, the estimation of its mass flux and of its connectivity to Earth.

There is now a broad consensus that heating in the lower solar atmosphere is an intermittent, multi-physics process that can hardly be captured without comprehensive 3D radiative-MHD (rMHD) models \citep[see e.g.,][]{carlssonNewViewSolar2019,przybylskiStructureDynamicsInternetwork2025,norazChromosphereQuietSun2026}. In addition to quantifying in details the small-scale dynamics (see Sect~\ref{sec:solar_chromo}), these models provide essential constraints to understand and parameterise its influence on mass and energy transfers to higher layers. This indeed will be key information for widely used approaches, such as 1-D field-aligned models \citep[e.g.][]{carlssonNonLTERadiatingAcoustic1992} or global coronal and solar-wind simulations in which the the low-atmospheric coupling is not explicitly resolved, but yet appears to be critical \citep[][]{parsonsWangSheeleyArgeEnlilConeModel2011,vanderholstALFVENWAVESOLAR2014,parentiValidationWaveHeated2022,perriCOCONUTNovelFastconverging2023}.

At the same time, these complex 3D rMHD models \citep[][]{gudiksenStellarAtmosphereSimulation2011,przybylskiChromosphericExtensionMURaM2022} still require observational constraints to be fully reliable. Yet the small-scale, rapidly evolving nature of shocks, current sheets, and linear-wave dynamics remains extremely challenging to observe, especially in quiet-Sun regions. The enhanced spatial resolution, polarimetric sensitivity, and temporal cadence of EST 
will be crucial for resolving these processes (cf. Sections \ref{sec:solar_chromo} \& \ref{sec:solar_coronaloscillations}), constraining advanced models, and improving our understanding of how lower-atmospheric physics couples to the corona and heliosphere, with direct implications for space-weather prediction.

Furthermore the chromospheric level of heating, wave activity, and turbulence has a direct impact on the transport of heavy ions into the corona and the solar wind (see Sects~\ref{sec:solar_helium_sw} and \ref{sec:solar_abundances}). These processes regulate the enrichment or depletion of different species in the upper atmosphere, thereby controlling the mass loading at the base of forecasting models. Most current space-weather prediction tools, however, still rely on single-fluid formulations, and fully multi-fluid treatments will likely remain out of operational reach for the foreseeable future \citep[][]{brchnelovaCOCONUTMFTwofluidIonneutral2023}.

In this context, advanced models that incorporate more complete coronal physics are essential. They enable the derivation of robust parameterisations for the behaviour of individual species, such as helium abundance or ion-specific fractionation, that can be directly incorporated into single-fluid operational frameworks. A more accurate characterisation of chromospheric heating and turbulence, made possible by high-resolution and high-sensitivity observations from facilities like EST, is therefore a key step toward improving the representation of the plasma at the base of the solar wind and, ultimately, strengthening the reliability of space-weather forecasting systems. 
Indeed, a better characterization of the solar wind will guaranty a better anticipation of fast-stream events, as well as a better description of the propagation of CMEs, which are in particular impacted by the velocity and density of the surrounding solar wind through the drag force \citep{Lavraud2015,Rouillard2020b}.

The final step is to be able to connect what we understand of the solar wind at the solar surface with what we measure near the Earth environment (in situ measurements). This is done through solar connectivity, which refers to which specific part of the Sun is the Earth connected, either through large-scale magnetic field or through solar wind flows. Indeed, most space weather events consist of electrically charged particles, which makes them very sensitive to magnetic field interactions. This information is however difficult to access, as measurements of the interplanetary magnetic field are currently only local at the position of certain probes, and not global all along the Sun-Earth axis. We know that due to the rotation of the Sun, they will not be radial, and are very likely to follow the large-scale Parker spiral, with local reconnections and distortions due to shear or waves \citep{Owens2013}. Another approach can be to use the abundances of heavier elements as described in section \ref{sec:solar_helium_sw}, but such methodology is for the moment not applicable in a systematic way \citep{Yardley2024}. However, it is a crucial constrain for space weather previsions, as it allows to focus on the most geo-effective part of the Sun to anticipate transient events.

Over the past few years, many efforts have been deployed by the french community to improve our understanding of both magnetic and solar wind connectivity. Since the knowledge of 3D magnetic field in the coronal volume cannot be readily obtained from observations, coronal magnetic field must be reconstructed (i.e. modelled), thanks to so-called magnetic extrapolation methods \citep[see e.g. review of][cf. the dedicated point in Section \ref{sec:spaceweather_flares}, see also Section \ref{sec:spaceweather_ambiguity}]{Wiegelmann17}, using observed 2D photospheric magnetograms. 
The connectivity tool described in \citet{Rouillard2020}, which is now part of the Solar Terrestrial ObseRvations and Modeling Service\footnote{ \url{https://storms-service.irap.omp.eu/}}
(STORMS), a SNO of INSU, 
consists in a PFSS (Potential-Field Source Surface) extrapolation based on photospheric magnetic field measurements with empirical hypothesis about the solar wind speed, and is used in real-time operations for missions such as SolO. Such a method was coupled to ballistic mapping in \citet{Bizien2025} to trace back PSP switchbacks to their solar origin. In \citet{Jarry2024} and \citet{Kennis2024}, magnetic tracing has also been applied to MHD simulations in order to model more complex magnetic field behaviors (such as reconnection, turbulence or waves).

However, studies point out the impact of spatial resolution on magnetic extrapolations, which could result in missing solar flux \citep{Arge2024}. Hence, having high-resolution images of the photospheric magnetic field, especially at the borders of active structures (such as active regions or coronal holes), thanks to EST observations, should help the french community improve their previsions of solar connectivity, and hence our understanding of the Sun-Earth connection for space weather previsions.


\subsection{Monitoring of active nests}
\label{sec:spaceweather_activenests}

Active regions (ARs) are magnetic regions of interest connected to many space weather phenomena (such as flares and CMEs). Therefore, they are actively monitored by many observatories. AR emergence is structure by the underlying dynamo, with magnetic flux as a function of latitude over the solar cycle (sunspot butterfly diagram) as well as clustering in Carrington longitude; almost 50\% of ARs appear close to another AR \citep{SchrijverZwaan2008}, in magnetically active zones called active nests that can survive several solar rotations. These magnetically active zones, or nests, are not systematically monitored yet, which makes it difficult to track them from one solar rotation to another, to understand how they form, evolve and disappear, and to link the ARs back to their original nest. However they carry an increased risk of solar flares/storms.

\begin{figure*}
    \centering
    \includegraphics[width=0.99\linewidth]{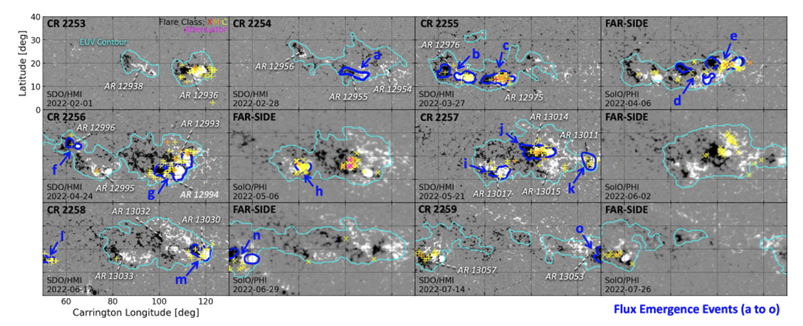}
    \caption{Snapshots of the photospheric magnetic field associated to an active nest from February to July 2022. We combine here multi-view measurements from SDO/HMI and SO/PHI. Solar flares associated are shown with crosses coloured by flare class. When available, the NOAA active region numbers are annotated. The arrows highlight flux emergence between each snapshot. From \cite{Finley2025}.}
    \label{fig:active_nest}
\end{figure*}

It has been especially difficult to monitor active nests due to the lack of 4$\pi$ observations of the Sun since the loss of STEREO-B, that lead us to miss the evolution of ARs on the far-side of the Sun \citep{Perri2024}. With the deployment of the SolO mission, it has become easier to track the active nests as the satellite can cover the far-side of the Sun for half a year. In \cite{Finley2025}, an active nest has thus been studied continuously for six months for the first time, showing that it contributed up to 70\% of the total solar flaring activity. \cite{Finley2024} has also shown that such active nests play a key role in structuring the solar wind for connectivity (see section \ref{sec:spaceweather_connectivity}). 

However, most of the available observations are large-field views of the solar disk. With the high-resolution images of the magnetic field that EST can provide thanks to its narrow-band spectropolarimeter combined with Stokes inversions, it would become possible to better understand the evolution of such active nests. Using the large-field monitoring to define the contours of the nest and provide pointing coordinates in advance, EST could focus on the most sensitive regions of the nests (such as active regions merging than can extend it or filament formation that can divide it) in order to help anticipate and understand the nest evolution. It would then be possible to monitor and even anticipate emergence of new flux in these regions, and see how the existing field gets destabilized. EST has the potential to create a database of long-lived active regions that would be preferential targets for future observations.

Since EST shall run for at least a full solar cycle, with such long-term programs, EST will be able to monitor the evolution of sunspots complexity in regards with the solar cycle, and thus provide key knowledge on the anticipation of rising activity on the Sun. The long term monitoring of active nest will permit to help better anticipate the regions of emergence of future ARs, and thus help improve space weather surveillance and forecasting.

\subsection{High-resolution characterisation of CME filaments}
\label{sec:spaceweather_filaments}

Coronal mass ejections (CMEs) are one of the most geo-effective space weather events that can trigger serious technological damages on Earth through geomagnetic storms \citep{Webb2012}. In order to anticipate their impact on Earth, several properties are key to predict: their trajectory of propagation, but also their orientation (southward-oriented CMEs are more efficient at penetrating the Earth magnetosphere due to the current Earth magnetic field geometry) and their magnetic budget \citep{Temmer2021}. However, predicting their impact is still a current scientific challenge: not only can CMEs' properties change during their propagation due to their interaction with the solar wind and interplanetary magnetic field \citep{Lavraud2014}, but we are also lacking 3D observations of their initial propagation phase close to the Sun. Hence, we have to rely heavily on interpretation of their initiation phase before their onset \citep{Verbeke2023}. This means that we need to have a better understanding of the filament structure which is leading to the CME eruption \citep{Schmieder2002}. 

Solar filaments (known as prominences at the solar limb) are relatively cold ($10^4$ K) and dense plasma embedded in the hot corona ($10^6$ K) \citep[cf. review of ]{Parenti14}. Filament fine structures are composed of thin and dark fibrils \citep{lin2005,Karki2025b}. All the filament fibrils are slightly inclined versus the polarity inversion line (PIL). In magnetic structures including magnetic dips, such as the flux rope (FR) and sheared arcade, the dense plasma is supported by the Lorentz force against the gravity force \citep{Aulanier1998,Aulanier1999,Aulanier2002, Guo2010}.

Regarding the link between filament onsets and CMEs, filament first starts to rise slowly \citep{Schmieder2002, chandra2021,devi2021}, then evolves to erupt through kink, torus instability and fast magnetic reconnection \citep{kliem2006, aulanier2010,xing2024}. Depending on the  environment of the FR, it can escape and accelerate particles in open magnetic field lines or be confined if the magnetic field strength is too strong above the FR \citep{Zuccarello16,Amari18}, or if the Lorentz force leads to an increase in the electric current, forcing the FR to rotate and ultimately stop its rising phase \citep{zhang2024}.  

\begin{figure*}[ht!]
    \sidecaption
    \includegraphics[width=12cm]{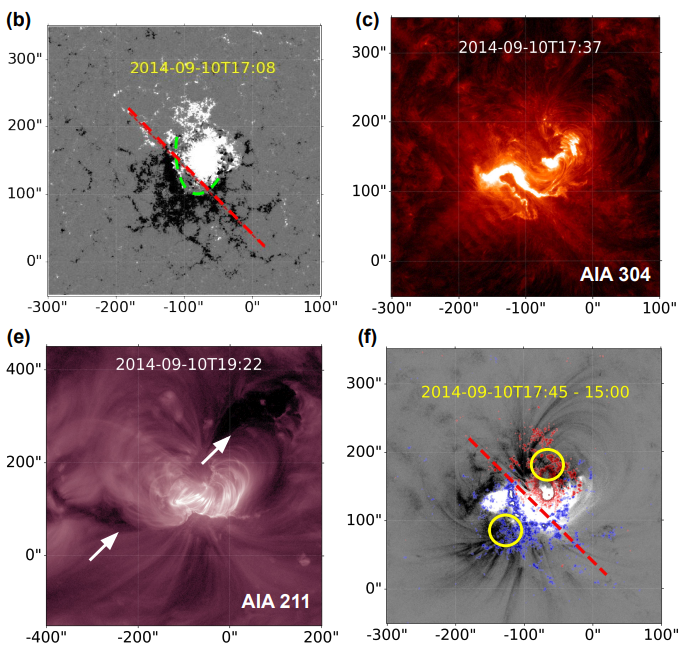}
    \caption{Snapshots of the multi-wavelength observation of an erupting filament in order to identify the corresponding CME parameters. Panel (b) shows the SDO/HMI magnetogram to identify the PIL and deduce the magnetic field inclination. Panel (c) shows a chromospheric image of SDO/AIA 304 to identify the shape of the flux rope and deduce its handedness. Panel (e) shows an EUV image by SDO/AIA 211 to identify the coronal dimmings associated with the CME release shock. Panel (f) shows the magnetogram with polarity to identify the feet of the flux rope. From \cite{Maharana2023}.}
    \label{fig:filament_cme}
\end{figure*}

There are various models for forming a filament, such as injection, levitation and/or evaporation-condensation (for this last model, also called TNE, see also Sec. \ref{sec:solar_coronalrain}). The injection model is more relevant for active region (AR) filaments, and the levitation model is relatively well accepted when chromosphere fibrils gather together and rise. Nevertheless, evaporation–condensation model is, to date, the best accepted. Many hydrodynamic or MHD simulations for the filament formation have been developed around this idea. MHD modeling  for filament as  flux rope has been intensively studied e.g. \citep{Guo2022,Schmieder2024}.
Extensive studies have been performed theoretically on the dips present in an FR to characterise parameters such as the dip depth, the dip extension along the main axis, and the number of dips \citep{Guo2022}.

The french community has become more and more involved in space weather CME modeling over the past years \citep{Aulanier13,Aulanier2019,Masson13,Masson2019,Perri22, perriCOCONUTNovelFastconverging2023}. 
For instance, the study of \cite{Regnault2023} has been able to insert a modified Titov-Démoulin flux rope \citep{Titov2014} inside a solar wind solution in order to study its propagation from the solar surface up to the Sun. At the STORMS data center, an analytical flux-rope model based on \cite{Chen1989} is used as a flux-rope propagation tool. The propagation of CME through the heliosphere until the Earth has been realized by coupling several MHD models for a theoretical flux rope \citep{Linan2025}. In order to improve these models, they need to be better constrained with observations.

This is where EST can help improve the french CME modelling efforts with high-resolution images of both the photospheric magnetic field below the CME and the associated filament in $\rm{H\alpha}$. The dynamics of filaments observed with spectroscopy and high resolution instruments could be a good indicator of the pre-eruptive state of the filament. Filaments are the sites of oscillations and counterstreamings in the pre-eruptive phase \citep{Aulanier2002, Schmieder2010, Karki2025b}. 
Small-scale dynamics of chromospheric plasmas ($\sim 100$ km.s$^{-1}$) have been associated with magnetic reconnection \citep{Karki2025a}. However, these measurements effectively yield spatially averaged velocities that represent only lower bounds on the true amplitudes due to limited resolution from THEMIS. Higher spatial resolution and time-series observations are required to disentangle the fine-scale dynamics and assess their impact on large-scale evolution, as will be possible with EST capacities.

As based on the works of \cite{Scolini2019,Maharana2023}, we can use filament observation to derive the propagation and orientation of a CME, and active regions magnetic field measurements to estimate the associated magnetic budget. With higher resolution, it will be possible to improve significantly the observational constraints, and thus the final prevision of the CME impact on our technological society.

\begin{figure*}[ht!]
    \centering
   \includegraphics[width = 0.95\linewidth]{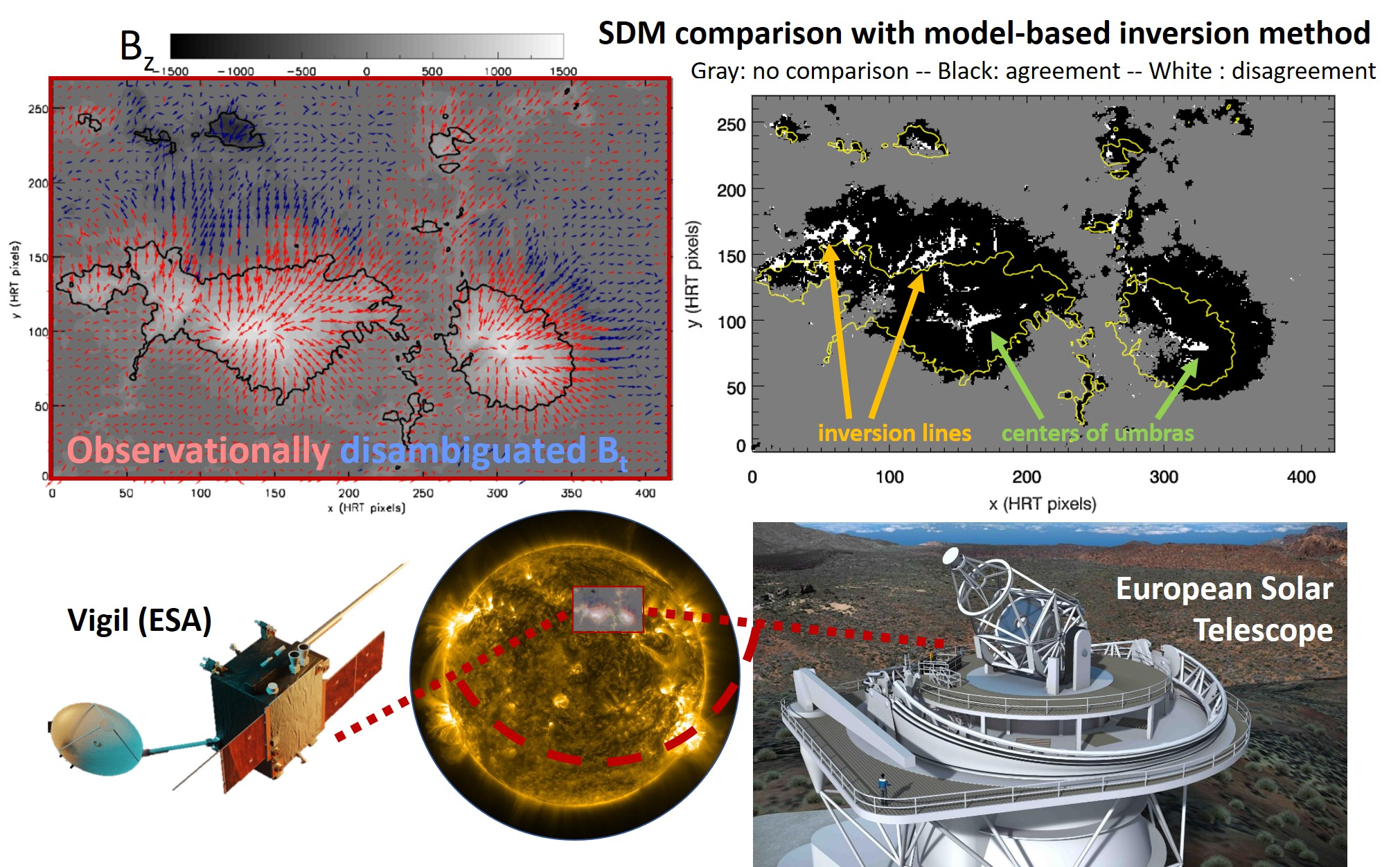}
    \caption{Upper left panel \citep[Adapted from][]{Valori23} : First fully-observational vector magnetic field map, with the ambiguity resolved thanks to the SDM method. Upper right panel \citep[Adapted from][]{Valori23}: comparison of the agreement on ambiguity of the transverse field of the SDM-disambiguated vector map with a vector map in which ambiguity is solved with a model-dependant method \citep[minimum energy method of][]{Metcalf94}. Main locations of disagreement are the core of the umbra and the inversion line. Bottom panel: concept of coordinated observation of to obtain vector magnetograms of active region from VIGIL with high-resolution high-quality reliable vector magnetic field measurements of region of interest (inversion line) by EST.}
    \label{fig:SDM_ESTVIGIL}
\end{figure*}

\subsection{Improving eruption prediction thanks to stereoscopic resolution of the transverse field 180° ambiguity} \label{sec:spaceweather_ambiguity}

The volume 3D structure of the coronal magnetic field is the key elements that controls the onset of solar eruptions \citep[e.g.][]{Aulanier10,Pariat17}. This coronal magnetic field cannot be directly estimated from observation and must be reconstructed from 2D photospheric magnetograms thanks to  extrapolations methods (cf. the dedicated topic in Section \ref{sec:spaceweather_flares}, see also Section \ref{sec:spaceweather_connectivity}). Unfortunately, all telescopes that performs spectropolarimetric measurements of the Zeeman effect have a blind spot: viewed from just one perspective, only the line-of-sight component of the magnetic field can be unambiguously determined, but not the direction of the magnetic field component perpendicular to it: this is the $180^\circ$ fundamental ambiguity. Therefore the magnetic field in its entirety cannot not be fully known from one  observation. To go beyond this knowledge gap reasonable assumptions about the unknown direction of the transverse magnetic field component are usually made \citep{Metcalf06}. However, those $180^\circ$ ambiguity resolution methods remains intrinsically model dependant. 

\citet{Valori22} proposed a different approach : the stereoscopic disambiguation method (SDM). Using two magnetographs observing the same target region from different vantage points, the measurement of the line-of-sight component from first magnetograph can be used to lift the orientation ambiguity of the transverse field measurement of the second magnetograph (and reversely). Taking advantages of the capacity of the Solar Orbiter mission with the PHI instrument on board to produce ambiguated vector magnetographs out of the Sun-Earth axis \citep{Rouillard20}, along with SDO/HMI magnetograms, \citet{Valori23} demonstrated that the SDM method could be effectively used to produced observationally-disambiguated maps (cf. Figure~\ref{fig:SDM_ESTVIGIL}, upper left panel).

 Comparison with standard model-dependant disambiguation method showed an overall good agreement but striking differences where noted by \citet{Valori23} on key locations: the center of umbras and magnetic inversion lines. The later are particularly interesting because that is precisely in these locations that strong electric current sheets form (cf. Section \ref{sec:solar_electriccurrents}). This indicates that standard model-dependant methods are likely missing key information at key location. As standard disambiguation method tends to choose "well-behaved" solutions, such as the minimum energy solution of \citep{Metcalf94,Metcalf06}, most of magnetograms disambiguated with model-dependant methods are likely incorrect in the most critical location where strong non-potential energy can be stored and where magnetic reconnection can take place. 
 
 \begin{figure*}[ht!]
    \centering
 \includegraphics[width=0.99\linewidth]{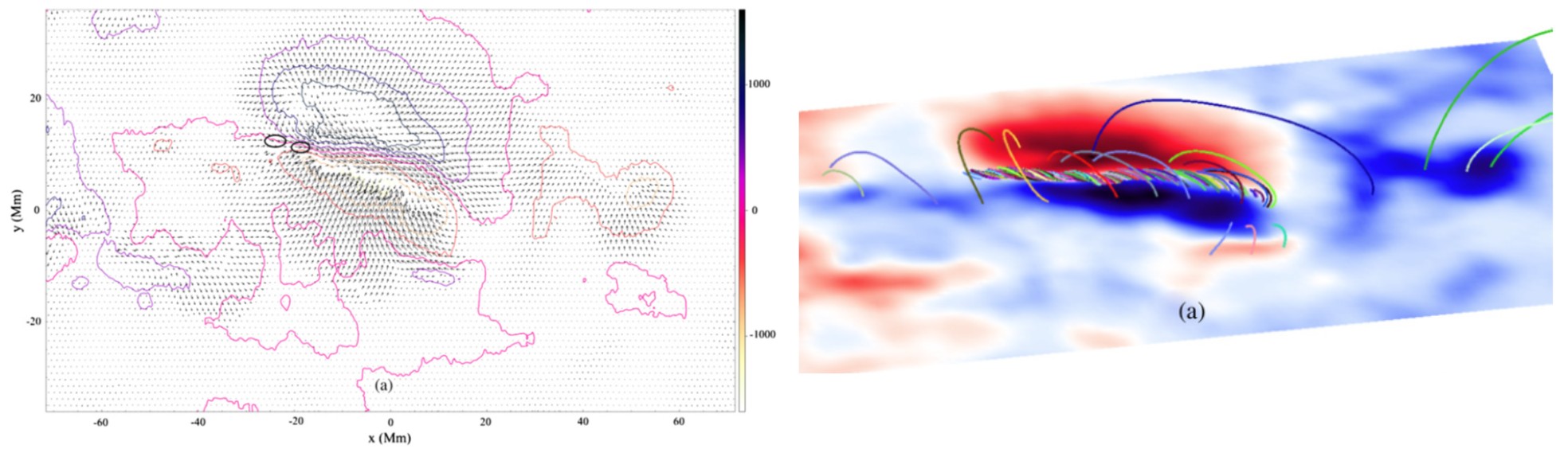}
    \caption{Example of a NLFFF magnetic extrapolation from a photospheric vector magnetogram. Left panel: Transverse magnetic field of active region NOAA 10808 measured by THEMIS on 2005 September 13 at 15:25 UT, with overlaid contour levels of the normal magnetic field (at $B_z=[0, \pm 400, \pm 800, and \pm 1200]$ G). Right panel: set of field lines reconstructed with the XTRAPOL as using the THEMIS vector magnetogram as boundary data, showing the presence of a twisted flux rope in equilibrium. Adapted from \citet{Canou09}.}
    \label{fig:extrapolation}
\end{figure*}
 
The 3D reconstruction models of active regions that are providing magnetic energy level and available energy for flares and eruptions, a central information for the space weather prediction of the impact of any solar eruption, are using vector magnetic field map as their key central input (cf. specific discussion on extrapolation in Section \ref{sec:spaceweather_flares}). Next generation of reconstruction methods of solar active regions are thus in dire need of proper vector magnetic field map. There is thus a key need of combined observation of:
\begin{itemize}
    \item Out of the Sun-Earth axis magnetograms  (line-of-sight is sufficient, although vector would be preferable). Such measurements can be obtain punctually by SolO/PHI but shall be readily and continuously available with the future ESA VIGIL mission. Data transfer and instrumental limitation on a spacecraft will however limit the spectral and spatial resolution of such observations
    \item High-fidelity measurements of magnetic fields and electric currents at key location of active regions, such as inversion line. This will be a key strength of EST observations (cf. Section \ref{sec:solar_electriccurrents}). Thanks to multi-line high-resolution observation with a high spectropolarimetric sensitivity, EST will provide the best magnetograms ever produced (still ambiguated). Despite its small field of view, by focusing on the location where its is known that non-potential energy can be stored and reconnection can favorably be induced, EST is perfectly suited to provide such high-fidelity measurements of magnetic field.
\end{itemize}

The association of both these measurements, carry the promise of ground breaking-improvements in 3D reconstruction of solar active region and thus strong improvement in space weather prediction of the threat level of observed active regions. French expertise in measurements of electric current density (cf. Section \ref{sec:solar_electriccurrents}), 3D reconstruction of coronal magnetic field (cf. Section \ref{sec:spaceweather_flares}), in disambiguation method \citep{Valori22,Valori23}, and with a strong expertise in observations-to-user approach thanks to its SNO (e.g. 3SOLEIL, CLIMSO, MEDOC,...) is ideally place to lead such developments.

\subsection{Anticipation of solar eruptions}
\label{sec:spaceweather_flares}

The question of the anticipation of solar eruptions is one of the biggest questions in space weather research. Indeed, the energetic particles accelerated during the release of energy, triggered thanks to the magnetic reconnection processes, travel close to the speed of light, which means that they reach Earth in at most 10 to 20 minutes. This short delay gives little to no time for an appropriate response from end-users to a real-time alert, meaning that it is essential to be able to anticipate solar eruptions before they occur.

Anticipation solely from observational data is still an open challenge. The EU-funded \href{http://flarecast.eu/}{Flarecast project} for example, in which several French teams were involved, tried to train neural networks and machine learning algorithms based on various available observations, and did not manage to find a reliable indicator for future flaring \citep{Georgoulis21}. Other innovative approaches are however keep being proposed, such as \citet{Barczynski20} that suggested that observational monitoring of the decay rate of surface currents could be used a proxy for CME acceleration when they are poorly observed between their onset and their appearance above the solar limb.

Another solution is to use data-constrained models of flux ropes, and to let them evolve to see if the flux rope tends to erupt or, on the contrary, to become stable again \citep[e.g.][]{Amari14,Amari18,Savcheva15,Savcheva16,Zuccarello16}. To be initiate, these models rely the knowledge of the magnetic field in the 3D coronal volume. Such information is not readily available from observations, although there are promising new developments in coronal magnetometry \citep[e.g.][]{Schad2024}. The coronal magnetic field must be reconstructed from measured observed data, most generally from observations of the distribution of the magnetic field at the solar photosphere, the so-called photospheric magnetograms (see Figure~\ref{fig:extrapolation}). Ideally vector magnetograms shall be obtained in which the 3 components of the magnetic field vector are determined observationally (with the limitation of the fundamental $180^\circ$ ambiguity, cf. \ref{sec:spaceweather_ambiguity}).

Using these observations as bottom boundary conditions, the rest of the magnetic field configuration is reconstructed, thanks to the so-called extrapolation methods \citep[e.g. see review of][]{Wiegelmann17}. These coronal-field reconstruction methods have been a core expertise of the French solar community the last 30-years, who have developed code using a variety of assumptions
The most common assumptions are the force-free (FF) ones, where one assumes, that the corona, where the plasma $\beta$ is very small, is quasi-relaxed at a magnetic quasi-equilibrium thanks to the Lorentz pseudo force, i.e. $\vec{j}\times\vec{B} =0$, implying that the electric current density, $\vec{j}$ is co-linear (and proportional) to the magnetic field. Among these FF approximation, we distinguish between the potential fields 
\citep[where $\vec{j}=0$, e.g.][]{Masson09}, 
the linear force free (LFF) fields \citep[$\vec{j}$ and $\vec{B}$ have the same constant of proportionality  e.g.][]{Aulanier00,Aulanier03,Schmieder03,Pariat04} 
and the non-linear force free (NLFF) fields \citep[where the level of proportionality varies field line by field line in the domain, e.g.][see example in Figure~\ref{fig:extrapolation}]{Regnier2004,Canou09,Amari13,Amari14,Amari25}. The most complete extrapolations method can go beyond this approximation and use non-force-free fields \citep[e.g. taking into account gravity, e.g.][]{Amari89,Aulanier1998} and/or derived from a complete numerical simulation.

\begin{figure*}
    \centering
 \includegraphics[width=0.99\linewidth]{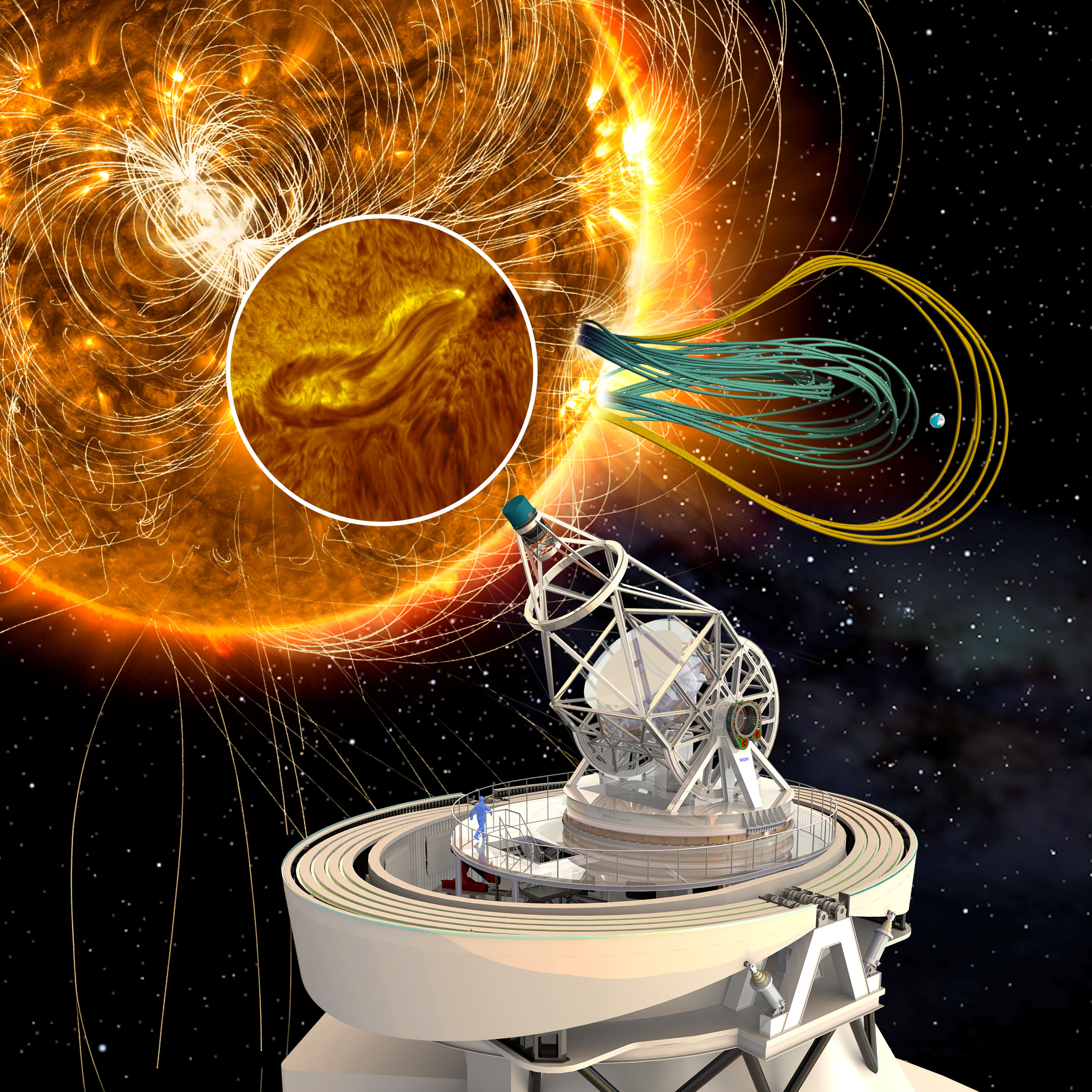}
    \caption{
    Illustrative image of how high-resolution observations obtained by EST (here HiFI/GREGOR H$_\alpha$ observation by Christoph Kuckein) can be used as inputs for numerical models of solar eruptions \citep[here adapted from]{Masson2019}. Advancing the understanding of solar eruptions, being a fundamentally multi-scale process, requires both data of the source regions at the lowest possible scale, and numerical models that can simulate the generation of the eruptive structures, track its propagation and determine its geoeffective-impacting properties.  Image composition produced by Gabriel Pérez Díaz (IAC).} 
  \label{fig:ESTspaceweather}
\end{figure*}

However, the accuracy of the extrapolation relies heavily on the accuracy of the magnetic field measurement at the photosphere. In section \ref{sec:solar_electriccurrents}, we already described a well-known measurement issue which is the difficulty to measure electric currents, and in section \ref{sec:spaceweather_ambiguity}, we described the issue of the $180^\circ$ ambiguity. Having high-resolution measurements from EST will significantly help improve these coronal magnetic field extrapolations, especially since it will provide a better description of the magnetic field at very low spatial scales. 
EST can also provide a unique opportunity to better adapt the post-treatment of such measurements, especially by assessing if the initial photospheric magnetic field is force-free or not (to justify which assumption can be applied in which case), and also to better address the flux balance done in most magnetograms (which can heavily affect the location of the magnetic field roots). 

In conclusion, high-resolution measurements of flux ropes embedded in active regions, such as those provided by EST, will significantly help the french community to improve the quality and truth-level of coronal magnetic field reconstruction, and thus the accuracy of flare and eruption previsions. This should lead to a breakthrough for the anticipation of the effect of flaring activity on Earth, such as radio blackouts and Global-Navigation-Satellite-System perturbations, as well as on the capacity to model the propagation and properties of CMEs as they impact the Earth, advancing forecasts reliability of the impact of geomagnetic storms on human technological activities.

%
%
%
%

\section{Conclusion: call for support and to a French participation to the EST project} \label{sec:conclusion}

While the 4-meter class Daniel K. Inoue Solar Telescope \citep[DKIST, ][]{Rimmele20} is now harvesting its first scientific results, at a time where India and China are respectively are planning the 2-meter Indian National Solar Telescope and the 8-meter Chinese Giant Solar Telescope, Europe presently only relies on several aging 1-m class national solar telescopes such as the Swedish Solar Telescope \citep{Scharmer2019}, the german GREGOR \citep{schmidt12}, and the French THEMIS  \citep{Gelly16,Schmieder25}, which have received their first light more than 15 years ago. 

Thanks to these facilities and their state-of-the-art instruments \citep[e.g. such as the swiss ZIMPOL spectropolarimeter][]{Gandorfer04}, Europe has presently the leading community in ground based solar physics and solar spectropolarimetry. Such expertise has and is still strongly sustaining the European expertise in both the astrophysics and space spectropolarimetry fields: major European institutes involved in solar ground-based instrumentation are also the main contributors to several space detectors, e.g. onboard SolO. Night-time astronomy has also always highly benefited from technology transfer from solar-physics spectropolarimetric developments  \citep[e.g. ZIMPOL/SPHERE on the Very Large Telescope][]{Schmid18}.

As of today, the 4-meter class European Solar Telescope \citep[EST, ][]{Noda22}, is the unique project in ground-based solar physics that will permit to maintain the present European leadership. EST shall become the most powerful European ground-based facility to study the Sun in the next 2-3 decades, in the visible and near-infrared bands. EST shall both complement and compete with DKIST. EST will be the leading facility for the quality of its spectropolarimetric sensitivity thanks to its on-axis design (unlike DKIST) and its state-of-the-art solar MCAO, and its next-generation instrumentation of spectral imagers and integral field units. It will also complement the time-coverage of observation and surveillance of the Sun due to DKIST and EST being almost at opposite longitudes on Earth. Very high spatial resolution observations, unreachable by space instruments, of space-weather-effect inducing regions will be monitored over a significantly larger period: e.g. regions driving space-weather effects will be monitored over extended periods, with very high spatial resolution observations, beyond current capabilities of space-based instruments.

The EST project is now at its ripest time for the effective start of its construction. The EST project has successfully passed major Preliminary Design Review (PDR) milestones, and a dedicated European Research Infrastructure Consortium (ERIC) is being consolidated with Spain, Slovakia and Czech Republic having committed to about 40\% of EST budget. National communities from other European countries are presently soliciting funds to their national agencies to complement the budget. Further delays would lead to a significant loss of pace in regards of the competing international projects and would weaken the European national heliophysics communities.

As discussed in Section \ref{sec:past}, since the start of the EST project, French researchers as well as high-tech French companies have demonstrated an interest in the EST project and tightly interacted with the European teams involved in the project. This interest has not dried out, and has on the contrary grown within the new generation of French  heliophysicists. The present review has highlighted and detailed 14 science cases that EST will tackle that are strongly in phase with the current investigations carried by French scientists, both in more "traditional" solar physics (cf. Section \ref{sec:solar}) and in innovative space weather research (see Section \ref{sec:spaceweather}). 

While ground-based solar instrumentation has been a strength of France in the second half of the XX\textsuperscript{th} century, fundamental research aiming at direct application toward space weather has become a leading theme and a strength of the French heliophysics community. France nationally-structure and service driven entities, such as OFRAME (cf. Introduction \ref{sec:Introduction}) and the "Services Nationaux d'Observations" (SNO, national observation services) of INSU, and most particularly those of the  "Actions Nationales pour l'Observation" (ANO, national observations action) "ANO-6 : Surveillance du Soleil et de l'environnement spatial de la Terre" (surveillance of the Sun and of the spatial environment of the Earth), e.g. 3SOLEIL, CDPP, CLIMSO, MEDOC, STORMS, etc., have ideally helped the consolidation of space weather oriented amenities. As section \ref{sec:spaceweather} illustrates, EST, by itself and in synergy with other facilities such as DKIST and VIGIL, will undoubtedly lift scientific locks that are presently dragging down the quality and precision of space weather forecasting. 

A new generation of French space-weather-aware researchers has been formed to the exploitation of high-resolution ground based instruments \citep[see e.g.][]{Froment2020,Joshi24,Joshi2025,NobregaSiverio24,Poirier2025} and further PhD students and postdocs are  presently being trained at their usage at diverse French laboratories, e.g. at THEMIS, LPP, LPC2E, ... France is thus ideally placed to exploit EST for space weather, strongly strengthening the expertise of the EST consortium on this aspect. 

French researchers working on the solar aspects of space weather, supported by numerous colleagues working at a vast numbers of French laboratories in heliophysics and astrophysics, namely 
CPHT, CRAL, DAp/IRFU, DPHY, IAS, IRAP, LAB, LAGRANGE, LATMOS, LIRA, LPC2E, LPP \& THEMIS
all co-authoring and co-signing this present white paper, agree that EST is a fundamental and necessary instrument for numerous space-weather oriented developments. Through this white paper, they urge the national community to participate to the funding of the EST project at a significant level. Not doing so, could impede the construction of this facility, leading to an irrecoverable loss of expertise at the European level in ground based solar physics and a missed opportunity at ground breaking development in space weather prediction, in particular for advanced prediction of solar eruptions. It must be noted that a French implication in EST would very likely imply an immediate return to investment for the French high-tech industrial network. High performance optical and opto-mechanical systems companies such as, e.g. ALPAO \& SAFRAN REOSC, which have been involved with the EST project throughout its development, would directly benefit from the EST construction. These industrial partners, which have been essential for the development of French astrophysics at large, being involved in numerous major project of the domain, are also supporting the EST project. The construction of the EST which would involved the polishing of a 4m telescope, could be critical in maintaining both the expertise and dedicated equipment at SAFRAN REOSC, being later ready for other next generation French astrophysics facilities. 

To conclude, the french participation to the EST project would thus absolutely not be at loss economically for France. It would be highly beneficial both for sustaining French industrial technological expertise and maintaining French leadership in space weather developments. For all the reasons detailed in this conclusion, supported by the strong scientific cases presented throughout this review, the co-signing authors call the French astrophysics community at large a support toward EST and ask to the French scientific administrative institutions to budget a French participation to the EST project.

\begin{acknowledgements}
The authors acknowledge the financial support of the French "Action Thématique Soleil-Terre" (ATST) of the Programme National ASTRO of CNRS/INSU, co-funded by CNES and CEA, to the Fédération de Recherche Plasmas à Paris - PLAS@PAR (FR2040), the logistic and administrative support of the Science and Engineering faculty of Sorbonne Université, as well as of the Laboratoire de Physique des Plasmas, that enabled the organisation of the "EST France 2025 Workshop", that is at the root of the present white paper. The authors thank the participants of the "EST France 2025 Workshop" for their insightful contributions and participation.  The contributing authors thank Sophie Masson as well as the EST Canarian Foundation for the help in the production of Figure~\ref{fig:ESTspaceweather}.
\end{acknowledgements}

\begin{funding}
     EP, NP, CF, GA are supported by the ATST of CNRS/INSU PN Astro, co-funded by CNES and CEA, and by financial support from the French national space agency (CNES) through the APR program. QN acknowledges funding support by the European Research Council (ERC) under the European Union’s Horizon 2020 research and innovation programme (grant agreement No 810218 WHOLESUN and No 101141362 Open SESAME) and by the Research Council of Norway through its Centres of Excellence scheme (RoCS project number 262622). CF and NP acknowledge funding by the Agence Nationale de la Recherche (ANR) for the CROSSWIND project under the grant ANR-24-CE31-2993. CF also benefited from the support of the International Space Science Institute (ISSI) ISSI Bern for International Team projects $\#401$ and $\#545$. 
\end{funding}
 


%
%


\bibliographystyle{aa.bst} 
\bibliography{ESTFrancewp.bib}



\end{document}